\documentclass{article}

\usepackage{amsmath}
\usepackage{amssymb}
\usepackage{float}
\usepackage{multirow}
\usepackage{lscape}
\usepackage{caption}
\usepackage{graphicx}
\usepackage{lipsum}
\usepackage{url}
\usepackage{dblfloatfix}
\usepackage{placeins}
\usepackage{arxiv}
\usepackage[utf8]{inputenc} 
\usepackage[T1]{fontenc}    
\usepackage{hyperref}       
\usepackage{url}            
\usepackage{booktabs}       
\usepackage{amsfonts}       
\usepackage{nicefrac}       
\usepackage{microtype}      
\usepackage{lipsum}

\usepackage{algorithm}
\usepackage{algorithmic}

\newcommand{\addcomment}[1]{{\color{black}{#1}}}

\usepackage{standalone}
\usepackage{tikz}
\usetikzlibrary{matrix}
\usetikzlibrary{arrows,positioning,shapes.geometric}
\usepackage{xcolor}
\colorlet{grey}{black!20}
\usepackage{soul}
\usepackage{hyperref}
\usepackage{acronym}

\newacro{ac}[AC]{alternating current}
\newacro{aes}[AES]{advanced encryption standard}
\newacro{avc}[AVC]{advanced video coding}

\newacro{B}[B]{bidirectional predicted}

\newacro{cabac}[CABAC]{context-adaptive binary arithmetic coding}
\newacro{cbr}[CBR]{constant bit rate}
\newacro{cfb}[CFB]{cipher feedback}
\newacro{cfe}[CFE]{call for evidence}
\newacro{cfp}[CFP]{call for proposal}
\newacro{ct}[CT]{constant time}
\newacro{ctc}[CTC]{common test condition}
\newacro{ctr}[CTR]{CounTeR}
\newacro{cpa}[CPA]{correlation power analysis}
\newacro{cpu}[CPU]{central processing unit}

\newacro{des}[DES]{data encryption standard}
\newacro{edr}[EDR]{edge differential ratio}
\newacro{egk}[EGk]{Exp-Golomb k-th order code} 
\newacro{eq}[EQ]{encryption qualilty}
\newacro{es}[ES]{encryption space}

\newacro{fl}[FL]{fixed length}

\newacro{hd}[HD]{hamming distance}
\newacro{hevc}[HEVC]{high efficiency video coding}
\newacro{hw}[HW]{hamming weight}
\newacro{I}[I]{Intra}
\newacro{iir}[IIR]{infinite impulse response}
\newacro{ipm}[IPM]{Intra prediction mode}
\newacro{iso}[ISO]{International Organization for Standardization}
\newacro{itu}[ITU]{International Telecommunication Union}
\newacro{itu-t}[\acs{itu}-T]{\acl{itu} Telecommunication Standardization Sector}
\newacro{iv}[IV]{initial vector}
\newacro{jctvc}[JCT-VC]{joint collaborative team on video coding}
\newacro{jpeg}[JPEG]{joint photographic experts group}
\newacro{jvet}[JVET]{joint video experts team}

\newacro{lfsr}[LFSR]{linear feedback shift register}
\newacro{lwcb}[LWCB]{light weight chaos-based}
\newacro{lsb}[LSB]{least significant bit}
\newacro{lut}[LUT]{lookup table}

\newacro{mpeg}[MPEG]{moving picture experts group}
\newacro{mppa}[MPPA\textregistered]{multi-purpose processing array}
\newacro{msb}[MSB]{most significant bits}
\newacro{mv}[MV]{motion vector}
\newacro{mvd}[MVD]{motion vector difference}

\newacro{ncpb}[NCpB]{number of cycles per byte}
\newacro{nist}[NIST]{national institute of standards and technology} 
\newacro{npcr}[NPCR]{number of pixels change rate}
\newacro{ntpb}[NTpB]{number of ticks per byte}
\newacro{ofb}[OFB]{output feedback}
\newacro{ott}[OTT]{over-the-top}
\newacro{P}[P]{predicted}
\newacro{pcng}[PCNG]{pseudo-chaotic number generator}
\newacro{prng}[PRNG]{pseudo-random number generator}
\newacro{psnr}[PSNR]{peak signal to noise ratio}
\newacro{pwlc}[PWLC]{piece wise linear chaotic}
\newacro{qp}[QP]{quantization parameter}
\newacro{qtc}[QTC]{quantized transform coefficient}

\newacro{ram}[RAM]{random-access memory}
\newacro{rc4}[RC4]{rivest Cipher 4}
\newacro{roi}[ROI]{region of interest}

\newacro{sao}[SAO]{sample adaptive offset}
\newacro{sca}[SCA]{side-channel attack}
\newacro{sc}[SC]{stream cipher}
\newacro{shvc}[SHVC]{scalable \ac{hevc}}
\newacro{ssim}[SSIM]{structural similarity}
\newacro{sss}[SSS]{Shamir’s secret sharing}
\newacro{sts}[STS]{statistical tests suite}
\newacro{sea}[SEA]{secure and embedded implementation}
\newacro{svc}[SVC]{scalable video coding}

\newacro{tb}[TB]{truncated binary}
\newacro{tsc}[TSC]{time stamp counter}
\newacro{tc}[TC]{transform coefficient}
\newacro{ts}[TS]{transform skip}
\newacro{trp}[TRp]{truncated rice code with context p}
\newacro{tucode}[TU]{truncated unary}
\newacro{uaci}[UACI]{unified average changing intensity}
\newacro{unary}[U]{unary}
\newacro{vliw}[VLIW]{very long instruction word}
\newacro{vmaf}[VMAF]{video multimethod assessment fusion}
\newacro{vceg}[VCEG]{video coding experts group}
\newacro{vtm}[VTM]{VVC test model}
\newacro{vvc}[VVC]{versatile video coding}

\newacro{xor}[XOR]{eXclusive OR}

\makeatletter
\usepackage[caption=false,font=normalsize, labelfont=sf, textfont=sf]{subfig}
\title{Selective Encryption of the Versatile Video Coding Standard }

\author{Guillaume Gautier,
        Mousa FarajAllah,
        Wassim Hamidouche,
        Olivier D\'eforges and
        Safwan El Assad
        \thanks{\color[rgb]{0.858,0.188,0.478}{The source code of the proposed method will be made publicly available.}}
\thanks{Guillaume Gautier, Wassim Hamidouche and Olivier D\'eforges are with INSA Rennes, Institut d'Electronique et des Technologies du num\'eRique (IETR), CNRS - UMR 6164, VAADER team, 20 Avenue des Buttes de Coesmes, 35708 Rennes, France (E-mails: firstname.lastname@insa-rennes.fr)}
\thanks{Mousa FarajAllah is with College of Information Technology and Computer Engineering, Palestine Polytechnic University, Palestine (Email : mousa\_math@ppu.edu)}
\thanks{Safwan El Assad is with Polytech Nantes, IETR, CNRS - UMR 6164, VAADER team (Email: safwan.elassad@univ-nantes.fr)}}

\begin{document}
\maketitle




\begin{abstract}
\Ac{vvc} is the next generation video coding standard developed by the \ac{jvet} 
\addcomment{and released} in July 2020. \Ac{vvc} introduces several new coding tools providing a significant coding gain over the \ac{hevc} standard. It is well known that increasing the coding efficiency adds more dependencies in the video bitstream making format-compliant encryption with the standard more challenging. In this paper we tackle the problem of selective encryption of the \ac{vvc} standard in format-compliant and constant bitrate. These two constraints ensure that the encrypted bitstream can be decoded by any \ac{vvc} decoder while the bitrate remains unchanged by the encryption. The selective encryption of all possible \ac{vvc} syntax elements is investigated. A new algorithm is proposed to encrypt in format-compliant and constant bitrate the \acp{tc} together with other syntax elements at the level of the entropy encoder. The proposed solution was integrated and assessed under the \ac{vvc} reference software model version 6.0. Experimental results showed that the encryption drastically decreases the video quality while the encryption is robust against several types of \addcomment{attacks}. The encryption space is estimated in the range of 15\% to 26\% of the bitstream size resulting in a lightweight encryption process. \addcomment{The web page of this work is available at \color[rgb]{0.858,0.188,0.478}{\url{https://gugautie.github.io/sevvc/}}}.                     
\end{abstract}


\markboth{Draft IEEE Transactions on Multimedia, February~2021}%
{Shell \MakeLowercase{\textit{et al.}}: Bare Demo of IEEEtran.cls for IEEE Journals}
\maketitle


\section{Introduction}

Security and confidentiality of multimedia contents are of prominent importance in many applications to ensure safe storage and transmission of images and videos. The straightforward solution to perform secure transmission of a video is to encrypt the whole video bitstream with a secure encryption protocol such as \ac{aes}~\cite{AES-FIPS}. However, this solution when applied to video has several limitations related to their high computational complexity increasing both the energy footprint and end-to-end latency. \addcomment{This increase in complexity/latency is mainly caused by the processing complexity of the encryption algorithm used to cipher the whole video especially when the video is encoded at high bitrate}. Moreover, the deciphering and ciphering processes are required to perform post-processing operations such as transcoding for network adaptation. This may harm security since the secret key is shared with \addcomment{untrusted middlebox in the network to perform splicing, quality monitoring, watermarking and transcoding}. The selective encryption solution has emerged as an effective alternative to perform secure and low complexity encryption of images and videos \cite{liu2006efficient}. The encryption process is \addcomment{performed} in the compressed-domain where only a set of the most sensitive information is encrypted. This enables performing both format-compliant and constant \addcomment{bitrate} encryption. The format-compliant property is very important enabling to decode the video bitstream without deciphering and thus all post-processing operations can be performed including packaging and transcoding without requiring access to the secret key used for encryption. Moreover, this property enables encrypting only some spatial regions in the image identified as \ac{roi} while keeping the rest of the image clear. The constant bitrate property preserves the encoder coding efficiency. Selective encryption has been widely investigated for different still image and video coding standards including \acs{jpeg}~\cite{jpegencry}, \acs{jpeg}-2000~\cite{JPEG200Survey, liu2006efficient}, \ac{avc}~\cite{5733402}, \ac{svc}~\cite{4624035} and more recently \ac{hevc}~\cite{shahid2013visual, farajallah2015roi, 7370952} and its scalable extension \acs{shvc}~\cite{hamidouche2017real}. 
Selective encryption of the \ac{hevc} standard has been widely investigated in the literature~\cite{van2013encryption, memos2016encryption, long2018format, sallam2018efficient} enabling format-compliant, secure and low complexity encryption. 


The \addcomment{\acs{iso}}/\ac{mpeg} and \addcomment{\acs{itu}}/\ac{vceg} developed the next generation video coding standard called \ac{vvc}. This latter, \addcomment{released in} July 2020, introduces new coding tools outperforming \ac{hevc} by up to 50\% in terms of bitrate reduction for a similar visual quality~\cite{8954562}. To the best of our knowledge, format-compliant and constant bitrate encryption of \ac{vvc} has not yet been addressed. Moreover, it is well known from information theory~\cite{ITShannon} that enhancing the coding efficiency adds more dependencies in the bitstream making format-compliant and constant bitrate encryption more challenging. 

This paper investigates a format-compliant and constant bitrate encryption of a video bitstream encoded with the \ac{vvc} standard. To meet these two constrains, the encryption is performed at the level of the \ac{cabac} engine. We first investigate all possible syntax elements that can be encrypted in both format-compliant and constant bitrate. A set of \ac{vvc} syntax elements including \ac{tc} values and signs, chroma prediction candidate, \ac{mv} differences and signs are encrypted. We propose a new algorithm that determines the encryptable bins within the \acp{tc}. The proposed selective encryption solution has been extensively assessed under the \ac{vvc} \acp{ctc} using three image and video quality assessment metrics including \ac{psnr}, \ac{ssim} and \ac{vmaf}, and security metrics such as \ac{eq}~\cite{EQmetric}, histogram analysis, edge detection and \ac{edr}~\cite{taneja2011chaos}. The proposed solution has also been tested against brute force attack, \ac{npcr} and \ac{uaci}~\cite{uacinpcr}. The encryption space giving the percentage of encrypted bits in the bitstream varies in the range of 15\% to 26\% for different targeted bitrates. This results in a very low decryption complexity which remains lower than 6\% of the decoding time.

The rest of this paper is organized as follows. Section~\ref{sec:selcenc} gives a brief review on selective encryption solutions proposed for \ac{hevc} and then Section~\ref{sec:cabac} describes the entropy coding of syntax elements in \ac{vvc}. Section~\ref{sec:solution} presents the proposed solution to encrypt \ac{vvc} syntax elements in format-compliant and constant bitrate. The performance of the selective encryption solution is assessed in Section~\ref{sec:results} in terms of video quality degradation, resilience to different attacks and complexity overhead. Finally, Section~\ref{sec:con} concludes this paper.


\section{Related works}
\label{sec:selcenc}
\begin{figure*}[!tbp]
\centering
\renewcommand{\arraystretch}{1.3} 
\resizebox{0.8\linewidth}{!}{    \begin{tikzpicture}[>=latex']
        \tikzset{block/.style= {draw, rectangle, align=center,minimum width=3.5cm,minimum height=1cm},}
        \node [block, fill=grey]  (SynElmt) {Syntax Elements};
        \node [block, right = 1cm of SynElmt, fill=grey]  (bin) {Binarization};
        
        \node [coordinate, right = 0.5cm of bin] (AR){};
        \node [coordinate, above = 1cm of AR] (AUR){};
        \node [coordinate, right = 0.5cm of bin] (BR){};
        \node [coordinate, below = 1cm of BR] (BBR){};
        
        \node [block, right = 1cm of AUR, fill=grey]  (CM) {Context Modeling};
        \node [block, right = 1cm of BBR, fill=green]  (SE) {Selective Encryption};
        
        \node [block, right = 1cm of CM, fill=grey]  (CC) {Context Coding};
        \node [block, right = 1cm of SE, fill=grey]  (BC) {Bypass Coding};
        
        \node [coordinate, right = 0.7cm of CC] (CR){};
        \node [coordinate, below = 1cm of CR] (CBR){};
        \node [coordinate, above = 1.25cm of CR] (CAR){};
        \node [coordinate, above = 0.75cm of CM] (AA){};
        \node [coordinate, right = 0.7cm of BC] (DR){};
        \node [coordinate, above = 1cm of DR] (DUR){};

        \node [block, right = 1cm of CBR, fill=grey]  (bitstream) {Bitstream};

        \node [coordinate, above right = 0.5cm of CC] (A){};
        \node [coordinate, above left  = 0.5cm of CC] (B){};
        \node [coordinate, below left  = 0.5cm of BC] (C){};
        \node [coordinate, below right = 0.5cm of BC] (D){};

        \path[draw, ->]
             (SynElmt) -- (bin);
        \path[draw, ->]
             (bin) -- (AR)
             (AR) -- (AUR)
             (AUR) -- (CM);
        \path[draw, ->]
             (bin) -- (BR)
             (BR) -- (BBR)
             (BBR) -- (SE);
        \path[draw, ->]
             (SE) -- (BC);
        \path[draw, ->]
             (CM) -- (CC);
        \path[draw, ->]
             (CC) -- (CR)
             (CR) -- (CBR)
             (CBR) -- (bitstream);
        \path[draw, ->]
             (BC) -- (DR)
             (DR) -- (DUR)
             (DUR) -- (bitstream);
        \path[draw, dashed, ->]
             (CC) -- (CR)
             (CR) -- (CAR) 
             (CAR) -- node[above]{Context update} (AA) 
             (AA) -- (CM);
        \path[draw, dash pattern={on 7pt off 3pt}]
             (A) -- (B) -- (C) -- node[below]{\textbf{Arithmetic Coding}} (D) -- (A);
    \end{tikzpicture}}
\caption{Overall architecture of the \ac{cabac} engine in \ac{vvc}. The selective encryption block is illustrated in green}
\label{CACABEngine}
\vspace{-3mm}
\end{figure*}
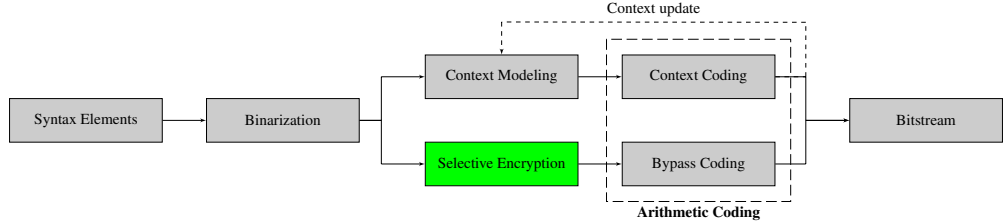

In this section, we review the existing solutions for \ac{hevc} standard encryption.
The first format-compliant encryption solution of \ac{hevc} was proposed by Shahid~{\it et al.}~\cite{shahid2013visual}. In this solution, \ac{aes} was used in \ac{cfb} mode to perform selective encryption of the selected syntax elements at the \ac{cabac} stage. This work considered an earlier version of the \ac{hevc} standard and some encryptable syntax elements of \ac{hevc} were not identified in this solution. Farajallah~{\it et al.}~\cite{farajallah2015roi} proposed a selective encryption solution to cipher the \ac{roi} in \ac{hevc} standard. This solution relies on the tile concept introduced in \ac{hevc} enabling a frame partitioning into independent rectangular regions. The encryption process encrypts only tiles within the \ac{roi} and keeps the background clear. The tiles within the \ac{roi} are encrypted in format-compliant and constant bitrate by ciphering only a set of sensitive syntax elements. Moreover, to prevent the encryption propagation outside the \ac{roi}, \acp{mv} of the background tiles are constrained to only refer to background area (no \ac{roi}) in the reference frames.  Boyadjis~{\it et al.}~\cite{7370952} presented a selective encryption algorithm in order to increase the visual distortion. The presented research moves selective encryption from bypass mode to regular mode, which negatively affects the bitrate. Luma intra prediction modes are selected to be encrypted in addition to the residuals. The presented solution enables more scrambling performance while the compression efficiency has changed leading to a slight bitrate increase. Hamidouche~{\it et al.}~\cite{hamidouche2017real} investigated a selective encryption of the final version of \ac{hevc}. The authors have proposed a real time selective encryption solution for the scalable extension of \ac{hevc} named \ac{shvc}. The presented solution has analyzed all \ac{shvc} syntax elements in order to perform format-compliant, constant bitrate and low latency encryption while preserving all \ac{shvc} features. The presented results showed the high security level of the selective encryption solution with a low complexity overhead below 6\% of the decoder complexity.
\addcomment{Van Wallendael~{\it et al.}~\cite{van2013encryption} presented a format-compliant selective encryption solution for \ac{hevc}. They selected a set of syntax elements from \ac{hevc} that preserve the format compliance. Several techniques to selectively encrypt the video are investigated. The obtained result showed that most of the selected syntax elements have a low effect on the rate-distortion performance while having a broad range in scrambling performance.} 
Memos~{\it et al.}~\cite{memos2016encryption} presented an algorithm that encrypts only \ac{I} frames of the \ac{hevc} bitstream based on the idea that \ac{P} and \ac{B} frames are useless without \ac{I} frame. Moreover, encrypting only \ac{I} frames will decrease the encryption time by 50\% and propagates the encryption to other frames. The presented algorithm  merged two algorithms proposed in~\cite{shamir1979share, vijayalakshmi2010efficient}, while introducing some modifications to the selection and management of the encrypted data to be amendable to \ac{hevc}. This work relies on the \ac{aes} algorithm for secure transmission of \ac{hevc} bitstream with 256 bits as key length. It collects sign bits of each transform coefficient of \ac{I} frames until the collected signs reached 256 bits. However, it is not clear in the proposed algorithm whether the collected bits are used as key value or as state value since \ac{aes}-256 state size is 128 and not 256. The proposed algorithm performs conventional \ac{aes} encryption on the collected bits and swap the  original sign bits by the encrypted ones. Finally, the \ac{sss} input parameters are collected from the non-zero \ac{ac} coefficients of each transform block within the \ac{I} frame.
It is clear from the description that the proposed algorithm performs partial encryption algorithm. It is also important to note that the proposed solution is not format-compliant, none constant bitrate since it increases the bitstream size at least by 8\%. 
Long~{\it et al.}~\cite{long2018format} presented a format-compliant encryption in order to secure \ac{hevc} streams in multimedia social networks. The presented algorithm is tightly integrated with the encoding/decoding processes. The presented work performs encryption in two steps. First, a stream cipher is used to encrypt sign of the nonzero \acp{tc}, and the first sign bit hiding of \acp{tc}. Second, based on a control factor, only one parameter from merging index, \ac{mv} prediction index, sign of \ac{mv} difference and reference frame index is encrypted. The presented research increases the bitrate, while it is format-compliant solution. Finally, the presented work was assessed regarding security and complexity which confirms the good security level and acceptable complexity overhead. 
Ahmed~{\it et al.}~\cite{sallam2018efficient} presented a new solution for efficient selective encryption based on the chaotic logistic map for \ac{hevc}. The presented solution encrypts the sign bit of the \ac{mv} differences and the \acp{tc}. The encryption process is performed at the entropy coding stage of the \ac{hevc} encoding process. They focused on achieving a low complexity ciphering targeting real time applications, constant bitrate and format-compliant encryption. The presented work was compared with the solution proposed in~\cite{van2013encryption} and the obtained results confirm the suitability for real time applications with an intermediate level of security.
Peng~{\it et al.}~\cite{peng2019tunable} presented a tunable selective encryption scheme for \ac{hevc} based on chroma \ac{ipm} and \acp{tc} scrambling. The presented work has two security levels. The first one encrypts \ac{hevc} syntax elements including Luma \acp{ipm}, Chroma \acp{ipm}, the suffix part of the \acp{tc}, sign and value of the \ac{mv} differences, merge index, advanced \ac{mv} prediction, reference frame index, and \ac{sao} filter parameters. The second security level relies on edge extraction of each transform block. The transform block coefficients are scrambled to increase the security level only when the current transform block contains edges. Finally, the \ac{aes} is used in \ac{ctr} mode in order to generate the pseudo-random number sequences. These sequences are used to encrypt all previously mentioned parameters with a simple \ac{xor} operation.
Xu~\cite{xudata} proposed to perform data hiding inside the selected encrypted bitstream of \ac{hevc}. The secret message is hidden using a \ac{qtc} modification technique. It only changes bits value based on the data hiding without changing the data size in bypass coding, which confirms that the obtained solution is constant bitrate and format-compliant. Since the used operation is a \ac{xor}, the extraction process of the hiding data can be achieved on both encrypted as well as original videos. Obtained results confirm the resilience of the presented work against replacement attacks. Moreover, the degradation on the video quality introduced by data hiding is negligible. However, the presented algorithm was not evaluated regarding important general video attacks such as \ac{uaci}, \ac{npcr}, \ac{edr}, \ac{eq}, histogram analysis and key sensitivity attacks.

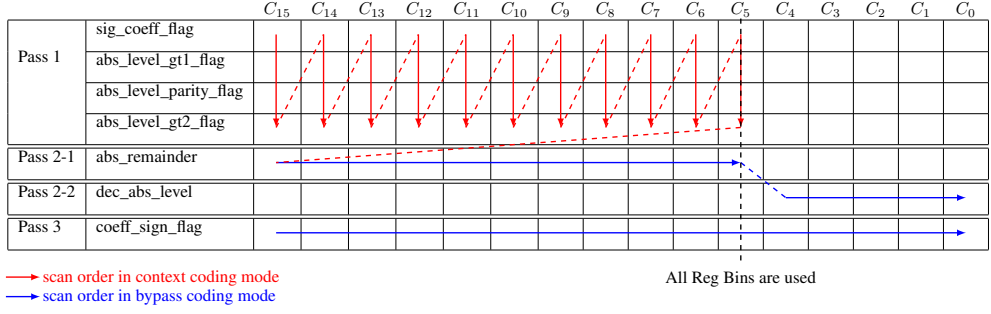
\begin{figure*}[tbp]
\centering
\resizebox{0.8\linewidth}{!}{\tikzset{>=latex}%
\tikzstyle{every picture}+=[remember picture, baseline]%
\tikzstyle{every node}+=[inner sep=0pt,anchor=base,
    minimum width=.5cm,align=center,text depth=.25ex,outer sep=1.5pt]%
\tikzstyle{every path}+=[very thick]

\newcommand{\tikzmark}[2]{\tikz \node (#1) {#2};}

\begin{tabular}{|l|l|p{0.5cm}|p{0.5cm}|p{0.5cm}|p{0.5cm}|p{0.5cm}|p{0.5cm}|p{0.5cm}|p{0.5cm}|p{0.5cm}|p{0.5cm}|p{0.5cm}|p{0.5cm}|p{0.5cm}|p{0.5cm}|p{0.5cm}|p{0.5cm}|}
\multicolumn{2}{c}{} & \multicolumn{1}{c}{$C_{15}$} & \multicolumn{1}{c}{$C_{14}$} & \multicolumn{1}{c}{$C_{13}$} & \multicolumn{1}{c}{$C_{12}$} & \multicolumn{1}{c}{$C_{11}$} & \multicolumn{1}{c}{$C_{10}$} & \multicolumn{1}{c}{$C_9$} & \multicolumn{1}{c}{$C_8$} & \multicolumn{1}{c}{$C_7$} & \multicolumn{1}{c}{$C_6$} & \multicolumn{1}{c}{\tikzmark{m-110}{$C_{5}$}} & \multicolumn{1}{c}{$C_{4}$} & \multicolumn{1}{c}{$C_{3}$} & \multicolumn{1}{c}{$C_{2}$} & \multicolumn{1}{c}{$C_{1}$} & \multicolumn{1}{c}{$C_{0}$} \\\hline
\multirow{4}{*}{Pass 1} & sig\_coeff\_flag & \tikzmark{m00}{} &\tikzmark{m01}{}&\tikzmark{m02}{}&\tikzmark{m03}{}&\tikzmark{m04}{}&\tikzmark{m05}{}&\tikzmark{m06}{}&\tikzmark{m07}{}&\tikzmark{m08}{}&\tikzmark{m09}{}&\tikzmark{m010}{}&\tikzmark{m011}{}&\tikzmark{m012}{}&\tikzmark{m013}{}&\tikzmark{m014}{}&\tikzmark{m015}{} \\[0.25cm] \cline{2-18} 
    & abs\_level\_gt1\_flag &&&&&&&&&&&&&&&&  \\[0.25cm] \cline{2-18}
    & abs\_level\_parity\_flag &&&&&&&&&&&&&&&& \\[0.25cm] \cline{2-18} 
    & abs\_level\_gt2\_flag & \tikzmark{m40}{}&\tikzmark{m41}{}&\tikzmark{m42}{}&\tikzmark{m43}{}&\tikzmark{m44}{}&\tikzmark{m45}{}&\tikzmark{m46}{}&\tikzmark{m47}{}&\tikzmark{m48}{}&\tikzmark{m49}{}&\tikzmark{m410}{}&\tikzmark{m411}{}&\tikzmark{m412}{}&\tikzmark{m413}{}&\tikzmark{m414}{}&\tikzmark{m415}{}\\[0.25cm]\hline\hline
Pass 2-1 & abs\_remainder &\tikzmark{m50}{}&\tikzmark{m51}{}&\tikzmark{m52}{}&\tikzmark{m53}{}&\tikzmark{m54}{}&\tikzmark{m55}{}&\tikzmark{m56}{}&\tikzmark{m57}{}&\tikzmark{m58}{}&\tikzmark{m59}{}&\tikzmark{m510}{}&\tikzmark{m511}{}&\tikzmark{m512}{}&\tikzmark{m513}{}&\tikzmark{m514}{}&\tikzmark{m515}{}\\[0.25cm] \hline\hline
Pass 2-2 & dec\_abs\_level &\tikzmark{m60}{}&\tikzmark{m61}{}&\tikzmark{m62}{}&\tikzmark{m63}{}&\tikzmark{m64}{}&\tikzmark{m65}{}&\tikzmark{m66}{}&\tikzmark{m67}{}&\tikzmark{m68}{}&\tikzmark{m69}{}&\tikzmark{m610}{}&\tikzmark{m611}{}&\tikzmark{m612}{}&\tikzmark{m613}{}&\tikzmark{m614}{}&\tikzmark{m615}{}\\[0.25cm]\hline\hline
Pass 3 & coeff\_sign\_flag &\tikzmark{m70}{}&\tikzmark{m71}{}&\tikzmark{m72}{}&\tikzmark{m73}{}&\tikzmark{m74}{}&\tikzmark{m75}{}&\tikzmark{m76}{}&\tikzmark{m77}{}&\tikzmark{m78}{}&\tikzmark{m79}{}&\tikzmark{m710}{}&\tikzmark{m711}{}&\tikzmark{m712}{}&\tikzmark{m713}{}&\tikzmark{m714}{}&\tikzmark{m715}{}\\[0.25cm]\hline

\multicolumn{10}{c}{}&\multicolumn{5}{c}{\tikzmark{m810}{}}&\multicolumn{3}{c}{}\\[0.01cm]
\multicolumn{10}{c}{}&\multicolumn{5}{c}{\tikzmark{m910}{All Reg Bins are used}}&\multicolumn{3}{c}{}\\[0.25cm]
\end{tabular}
\begin{tikzpicture}[overlay,remember picture]

\draw [->][red,thick] {(m00.center)  -- (m40.center)};
\draw [->][red,thick] {(m01.center)  -- (m41.center)};
\draw [->][red,thick] {(m02.center)  -- (m42.center)};
\draw [->][red,thick] {(m03.center)  -- (m43.center)};
\draw [->][red,thick] {(m04.center)  -- (m44.center)};
\draw [->][red,thick] {(m05.center)  -- (m45.center)};
\draw [->][red,thick] {(m06.center)  -- (m46.center)};
\draw [->][red,thick] {(m07.center)  -- (m47.center)};
\draw [->][red,thick] {(m08.center)  -- (m48.center)};
\draw [->][red,thick] {(m09.center)  -- (m49.center)};
\draw [->][red,thick] {(m010.center)  -- (m410.center)};

\draw [->][blue,thick] {(m50.center)  -- (m510.center)};
\draw [->][blue,thick] {(m611.center)  -- (m615.center)};
\draw [->][blue,thick] {(m70.center)  -- (m715.center)};

\draw [dashed][red,thick] {(m40.center)  -- (m01.center)};
\draw [dashed][red,thick] {(m41.center)  -- (m02.center)};
\draw [dashed][red,thick] {(m42.center)  -- (m03.center)};
\draw [dashed][red,thick] {(m43.center)  -- (m04.center)};
\draw [dashed][red,thick] {(m44.center)  -- (m05.center)};
\draw [dashed][red,thick] {(m45.center)  -- (m06.center)};
\draw [dashed][red,thick] {(m46.center)  -- (m07.center)};
\draw [dashed][red,thick] {(m47.center)  -- (m08.center)};
\draw [dashed][red,thick] {(m48.center)  -- (m09.center)};
\draw [dashed][red,thick] {(m49.center)  -- (m010.center)};

\draw [dashed][red,thick] {(m410.center)  -- (m50.center)};
\draw [dashed][blue,thick] {(m510.center)  -- (m611.center)};
\draw [dashed][black,thick] {(m-110.south)  -- (m810.center)};

\draw (-19.5,-2.5) coordinate (CC);
\draw (-19.5,-2.9) coordinate (BC);
\draw [->][red,thick] {(-20.2,-2.5)  -- (CC)node[right]{scan order in context coding mode} } ;
\draw [->][blue,thick] {(-20.2,-2.9)  -- (BC) node[right]{scan order in bypass coding mode} } ;

\end{tikzpicture}}
\caption{Binarization of the \acfp{tc} of a 4$\times$4 sub-block in \ac{tc} mode~JVET-S2002~\cite{jvetVTM6doc}.}
\label{TCsCoding}
\vspace{-3mm}
\end{figure*}

\section{\sc{CABAC engine in \ac{vvc}}}
\label{sec:cabac}
The \ac{cabac} engine defined in \ac{vvc} is similar to \ac{hevc} consisting of three main functions: binarization, context modeling and arithmetic coding~\cite{6317157}. The overall \ac{cabac} architecture is illustrated in Fig.~\ref{CACABEngine}. First, the binarization step converts syntax elements to binary symbols (bins). Second, the context modeling updates the probabilities of bins, and finally the arithmetic coding compresses the bins into bits according to the estimated probabilities.
\subsection{binarization methods}
\label{sec:bincabac}
Six binarization methods are used in \ac{vvc}, namely \ac{unary}, \ac{tucode}, \ac{fl}, \ac{tb}, \ac{trp} and \ac{egk}. The \ac{unary} code represents an unsigned integer \addcomment{$B$} with a binstring of length \addcomment{$B+1$} composed of \addcomment{$B$} 1-bins followed by one 0-bin. The \ac{tucode} code is defined with the largest possible value of the syntax element $cMax$ \addcomment{($0 \leq B \leq cMax$)}. When the syntax element value \addcomment{$B < cMax$}, the \ac{tucode} is equivalent to \ac{unary} code, otherwise \addcomment{$B$} is represented by a binstring of $cMax$ 1-bins.
The \ac{fl} code represents a syntax element \addcomment{$B$}  with its binary representation of length $\lceil \log_2(cMax + 1) \rceil$ with $\lceil x \rceil$ is smallest integer greater than or equal to $x$.
The \ac{tb} code is similar to the \ac{fl} code, except when the $cMax + 1$ value is not a power of 2. In this case, let $k$ be $k = \lfloor \log_2(cMax + 1) \rfloor$ (with $\lfloor x \rfloor$ is largest integer less than or equal to $x$). The first $u = 2^{k+1} - cMax$ elements are coded with a \ac{fl} code of length $k$. The remaining $cMax+1-u$ symbols are offseted by $u$ and coded by $k+1$ bins. The \ac{trp} code is a concatenation of a quotient \addcomment{$q = \lfloor B/2^p \rfloor$} and a remainder \addcomment{$r = B-q2^p$}. The quotient $q$ is first represented by the \ac{tucode} code as a prefix concatenated with a suffix $r$ represented by the \ac{fl} code of length $p$. The \ac{egk} code is also a concatenation of prefix and suffix. The prefix part of the \ac{egk} code is the \ac{unary} representation of \addcomment{$l(B)= \lfloor \log_2(\frac{B}{2^k}+1) \rfloor$}. The suffix part is the \ac{fl} code of \addcomment{$B+2^k(1-2^{l(B)})$ with $cMax=k+l(B)$}.
\subsection{\sc{\acfp{tc} coding}}
In this section we describe the \ac{cabac} coding of the \acp{tc}. Similar to \ac{hevc}, \ac{vvc} coefficients are either coded in regular \acp{tc} mode or \ac{ts} mode. In both modes, the transform block is first divided into sub-blocks.  
\subsubsection{\sc{\ac{vvc} \ac{tc} coding mode}}
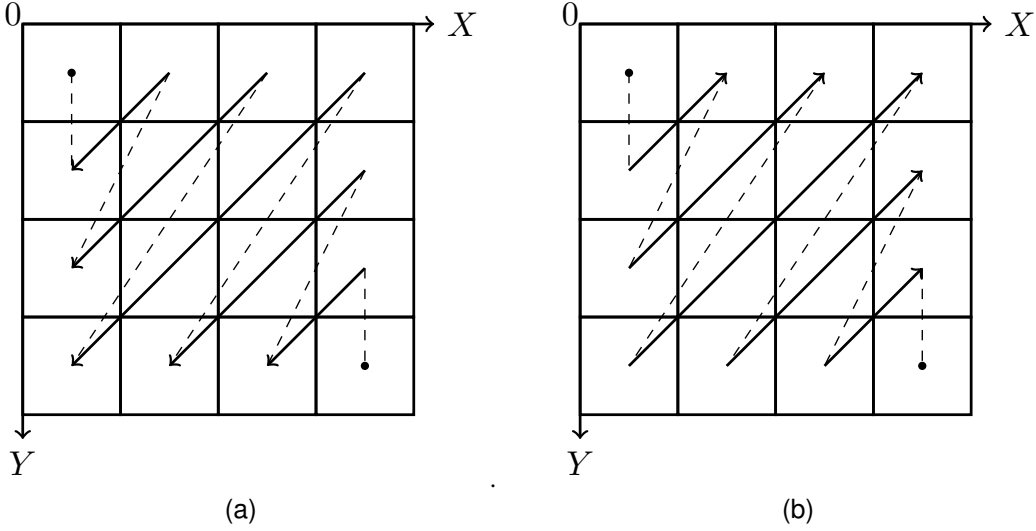
\begin{figure}[ht]
    \centering
    \subfloat[\label{TCsCourse}]{\resizebox{0.4\linewidth}{!}{\begin{tikzpicture}[thick]

\tikzset{square matrix/.style={
    matrix of nodes,
    column sep=-\pgflinewidth, row sep=-\pgflinewidth,
    nodes={draw,
      minimum height=#1,
      anchor=center,
      text width=#1,
      align=center,
      text=white,
      inner sep=0pt
    },
  },
  square matrix/.default= 1cm
}

\matrix[square matrix] at (1.5,-1.5)
{
$C_{0}$ & $C_{2}$  & $C_{5}$  & $C_{9}$  \\
$C_{1}$ & $C_{4}$  & $C_{8}$  & $C_{12}$ \\
$C_{3}$ & $C_{7}$  & $C_{11}$ & $C_{14}$ \\
$C_{6}$ & $C_{10}$ & $C_{13}$ & $C_{15}$ \\
};

\coordinate (0) at (0 ,-0) {};
\coordinate (1) at (0 ,-1) {};
\coordinate (2) at (1 ,-0) {};
\coordinate (3) at (0 ,-2) {};
\coordinate (4) at (1 ,-1) {};
\coordinate (5) at (2 ,-0) {};
\coordinate (6) at (0 ,-3) {};
\coordinate (7) at (1 ,-2) {};
\coordinate (8) at (2 ,-1) {};
\coordinate (9) at (3 ,-0) {};
\coordinate (10) at (1 ,-3) {};
\coordinate (11) at (2 ,-2) {};
\coordinate (12) at (3 ,-1) {};
\coordinate (13) at (2 ,-3) {};
\coordinate (14) at (3 ,-2) {};
\coordinate (15) at (3 ,-3) {};

\draw [->] {(14)  -- (13)};
\draw [->] {(12)  -- (10)};
\draw [->] {(9)  -- (6)};
\draw [->] {(5)  -- (3)};
\draw [->] {(2)  -- (1)};

\draw [fill=black,thin] (0) circle (1pt);
\draw [fill=black,thin] (15) circle (1pt);

\draw [dashed,thin] {(14)  -- (15)};
\draw [dashed,thin] {(12)  -- (13)};
\draw [dashed,thin] {(9)  -- (10)};
\draw [dashed,thin] {(5)  -- (6)};
\draw [dashed,thin] {(2)  -- (3)};
\draw [dashed,thin] {(0)  -- (1)};

\node (0) at (-0.6,0.6) {$0$};
\coordinate (O) at (-0.5,0.5) {};
\node (X) at (4,0.5) {$X$};
\node (Y) at (-0.5,-4) {$Y$};
\draw [->] {(O)  -- (X)};
\draw [->] {(O)  -- (Y)};

\end{tikzpicture}}}. \hspace{5mm}
    \subfloat[\label{TCsCourseTS}]{\resizebox{0.4\linewidth}{!}{\begin{tikzpicture}[thick]

\tikzset{square matrix/.style={
    matrix of nodes,
    column sep=-\pgflinewidth, row sep=-\pgflinewidth,
    nodes={draw,
      minimum height=#1,
      anchor=center,
      text width=#1,
      align=center,
      text=white,
      inner sep=0pt
    },
  },
  square matrix/.default= 1cm
}

\matrix[square matrix] at (1.5,-1.5)
{
$C_{0}$ & $C_{2}$  & $C_{5}$  & $C_{9}$  \\
$C_{1}$ & $C_{4}$  & $C_{8}$  & $C_{12}$ \\
$C_{3}$ & $C_{7}$  & $C_{11}$ & $C_{14}$ \\
$C_{6}$ & $C_{10}$ & $C_{13}$ & $C_{15}$ \\
};

\coordinate (0) at (0 ,-0) {};
\coordinate (1) at (0 ,-1) {};
\coordinate (2) at (1 ,-0) {};
\coordinate (3) at (0 ,-2) {};
\coordinate (4) at (1 ,-1) {};
\coordinate (5) at (2 ,-0) {};
\coordinate (6) at (0 ,-3) {};
\coordinate (7) at (1 ,-2) {};
\coordinate (8) at (2 ,-1) {};
\coordinate (9) at (3 ,-0) {};
\coordinate (10) at (1 ,-3) {};
\coordinate (11) at (2 ,-2) {};
\coordinate (12) at (3 ,-1) {};
\coordinate (13) at (2 ,-3) {};
\coordinate (14) at (3 ,-2) {};
\coordinate (15) at (3 ,-3) {};

\draw [<-] {(14)  -- (13)};
\draw [<-] {(12)  -- (10)};
\draw [<-] {(9)  -- (6)};
\draw [<-] {(5)  -- (3)};
\draw [<-] {(2)  -- (1)};

\draw [fill=black,thin] (0) circle (1pt);
\draw [fill=black,thin] (15) circle (1pt);

\draw [dashed,thin] {(14)  -- (15)};
\draw [dashed,thin] {(12)  -- (13)};
\draw [dashed,thin] {(9)  -- (10)};
\draw [dashed,thin] {(5)  -- (6)};
\draw [dashed,thin] {(2)  -- (3)};
\draw [dashed,thin] {(0)  -- (1)};

\node (0) at (-0.6,0.6) {$0$};
\coordinate (O) at (-0.5,0.5) {};
\node (X) at (4,0.5) {$X$};
\node (Y) at (-0.5,-4) {$Y$};
\draw [->] {(O)  -- (X)};
\draw [->] {(O)  -- (Y)};

\end{tikzpicture}}}
        \centering
        \caption{\acfp{tc} scanning orders (\ref{TCsCourse}) reverse diagonal scan order and~(\ref{TCsCourseTS}) regular diagonal scan order.}
    \label{fig:course}
\end{figure}

The coefficients of each sub-block are encoded in three passes as illustrated in Fig.~\ref{TCsCoding} for a 4$\times$4 sub-block. The coefficients are processed in reverse diagonal scan order, as depicted in Fig.~\ref{TCsCourse}. The first pass processes a group of flags until it reaches a limit of used bins specified by the standard. \addcomment{This maximum number of bins used in the first pass is computed with respect to the block size ($W_b \times  H_b$)  as follows $\lfloor (2^{\log_2 (W_b) + \log_2 (H_b)})\,7 / 4 \rfloor $.}
Once this limit is reached, the second pass starts encoding the remainders computed from the coefficient value $C$ as follows
\begin{equation}
abs\_remainder =  \left\{\begin{array}{lr}
\left \lfloor  \frac{| C | - 4}{2} \right \rfloor & | C | \geq 4, \\
0 & \text{otherwise}. 
\end{array}\right.
\end{equation}
\addcomment{The coefficients of value lower than 4 are binarized by the flags in the first pass.}
The second pass relies on the TRp/EGk binarization, until the position of the last coefficient processed by the first pass is reached. Then, the $dec\_abs\_level$ syntax element, computed by~\eqref{eq2:decabs} for the remaining coefficients is bypassed and binarized also using the TRp/EGk binarization. 
\begin{equation}
dec\_abs\_level = \left\{\begin{array}{lr}
        V, & \text{if } C  = 0,\\
        | C |, & \text{if } | C | \leq V,\\
        | C | + 1, & \text{if } | C | > V,
        \end{array}\right .
        \label{eq2:decabs}
\end{equation}
\addcomment{the $V$ constant is derived from a \ac{lut} $VArr$ according to the state and the local absolute sum $LocAbsSum$ computed for the current coefficient by~\eqref{eq:localabssum}. The $V$ value updates the 0 coefficient value such that coefficients have smaller binarization when large coefficients are mixed with definite 0 values}. The $LocAbsSum$ is a saturated sum in the interval $[0,31]$ of a set of neighboring coefficients $S_1$ illustrated in green in Fig.~\ref{TCsSum}
\begin{equation}
    LocAbsSum = \left [\sum_{i \in S_1} |C_i| - 5 \, BaseLvl \right ]^{31}_0,
    \label{eq:localabssum}
\end{equation}
where $BaseLvl$ is equal to 4 for the $abs\_remainder$ (Pass 2-1) and 0 for the $dec\_abs\_level$ (Pass 2-2). 

Finally, the third pass encodes the signs of the coefficients. \addcomment{We can notice that only the first pass relies on \ac{cabac} context coding and the last two passes perform bypass coding}. The {\it abs\_remainder} and {\it dec\_abs\_level} syntax elements are both binarized by a combination of TRp and in a special case EGk code. This binarization is presented in Section~\ref{subsec:bin}.


\subsubsection{\sc{\ac{vvc} \ac{ts} coding mode}}
In \acl{ts} mode, the coefficients of each sub-block are also encoded in three passes that process the coefficients in a simple diagonal scan order, as shown in Fig.~\ref{TCsCourseTS}. The first pass mainly encodes all coefficients considered as significant (ie. $ C  \neq 0$) including its sign and parity. The second pass encodes more flags to check whether the coefficient is greater than a certain threshold.  Finally, the third pass encodes the remainder coefficients greater than 10 using the TRp/EGk binarization of $abs\_remainderTS$
\begin{equation}
abs\_remainderTS = \left \lfloor  \frac{| C | - 10}{2} \right \rfloor.    
\end{equation}

The local absolute sum in the case of \ac{ts} mode $LocAbsSumTS$ is computed by \eqref{eq:localabssumTS} as follows
\begin{equation}
    LocAbsSumTS = \left [\sum_{i \in S_2} |C_i| \right ]^{31}_0.
    \label{eq:localabssumTS}
\end{equation}
It should be noted that the third pass relies on bypass coding. 

\subsubsection{\sc{Binarization process}}
\label{subsec:bin}
\begin{figure}[ht]
\centering
\vspace{-3em}
    \subfloat[\label{TCsSum}
]{\resizebox{0.4\linewidth}{!}{\begin{tikzpicture}[thick]

\tikzset{square matrix/.style={
    matrix of nodes,
    column sep=-\pgflinewidth, row sep=-\pgflinewidth,
    nodes={draw,
      minimum height=#1,
      anchor=center,
      text width=#1,
      align=center,
      inner sep=0pt
    },
  },
  square matrix/.default= 1.2cm
}

\matrix[square matrix]
{
|[fill=yellow]|$\ \ \ X_c\ \ \ $ $Y_c$  & |[fill=green]|$X_c+1$ $Y_c$  & |[fill=green]| $X_c+2$ $Y_c$ 		   \\
|[fill=green]|$\ \ \ X_c\ \ \ $ $Y_c+1$  & |[fill=green]| $X_c+1$ $Y_c+1$ &  \\
|[fill=green]|$\ \ \ X_c\ \ \ $ $Y_c+2$  &     &  \\
};
\coordinate (O) at (-1.8,1.8) {};
\node (X) at (2.3,1.8) {$X$};
\node (Y) at (-1.8,-2.3) {$Y$};
\draw [->] {(O)  -- (X)};
\draw [->] {(O)  -- (Y)};
\end{tikzpicture}}} \hspace{5mm}
    \subfloat[\label{TCsSumTS}
]{\resizebox{0.4\linewidth}{!}{\begin{tikzpicture}[thick]

\tikzset{square matrix/.style={
    matrix of nodes,
    column sep=-\pgflinewidth, row sep=-\pgflinewidth,
    nodes={draw,
      minimum height=#1,
      anchor=center,
      text width=#1,
      align=center,
      inner sep=0pt
    },
  },
  square matrix/.default= 1.2cm
}

\matrix[square matrix]
{
                                         & |[fill=green]|$X_c$ $Y_c-1$ \\
|[fill=green]|$\ X_c-1\ $ $Y_c$  & |[fill=yellow]|$\ \ \ X_c\ \ \ $ $Y_c$  \\
};
\coordinate (O) at (-1.2,1.2) {};
\node (X) at (2.3,1.2) {$X$};
\node (Y) at (-1.2,-2.3) {$Y$};
\draw [->] {(O)  -- (X)};
\draw [->] {(O)  -- (Y)};
\end{tikzpicture}}}
\caption{\acp{tc} dependencies: coefficients highlighted in green are used to compute the local absolute sum of the current coefficient in yellow for (\ref{TCsSum}) \ac{tc} mode and (\ref{TCsSumTS}) \ac{ts} mode. $S_1$ and $S_2$ are two sets of green coefficients in (\ref{TCsSum}) and (\ref{TCsSumTS}), respectively.}
\label{TCsSumAll}
\end{figure}
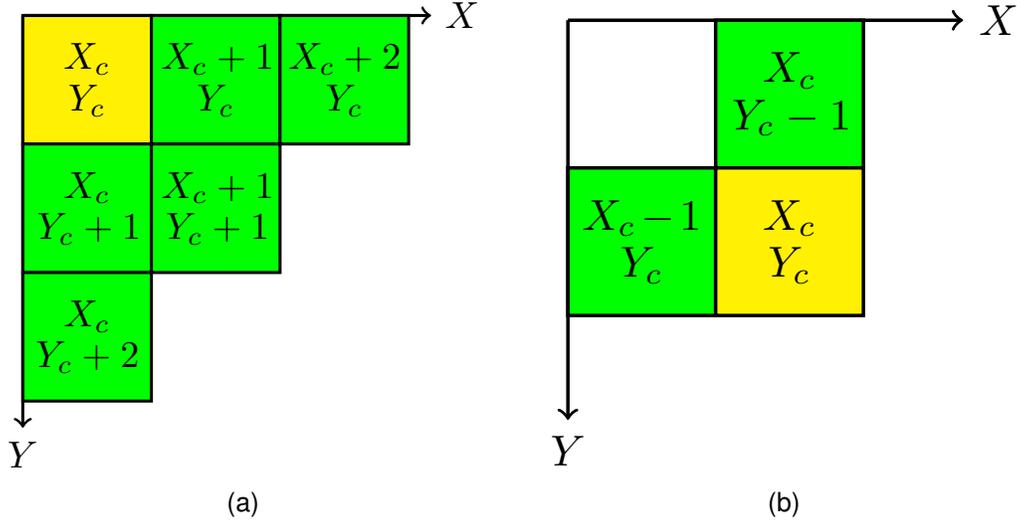

Algorithm~\ref{TcBinariation} gives the binarization process of the $abs{\_}remainder$. The $dec{\_}abs{\_}level$ and $abs{\_}remainderTS$ syntax elements are also binarized by this algorithm.
\addcomment{
The \acp{tc} remainders are binarized using either a \ac{trp} code, introduced in Section~\ref{sec:bincabac}, or an \ac{egk} code limiting the maximal length of a binarization to 32 bits as presented in Algorithm~\ref{limitedEGk}. The selection between the two binarizations depends on a threshold value $\beta$  defined in the standard as
\begin{equation}
    \beta = BinReduc \; 2^{cRiceParam},
\end{equation}
where $BinReduc$ is set to 5, and $cRiceParam \in \{ 0, 1, 2, 3\}$ is the rice parameter derived from a \ac{lut} $riceArr$ according to the saturated local absolute sum $LocSumAbs$ of previously coded coefficients computed by~\eqref{eq:localabssum}.} Fig.~\ref{TCsSum} illustrates in green the set of coefficients $S_1$ used to derive the rice parameter of the current coefficient highlighted in yellow. \addcomment{Similarly, Fig.~\ref{TCsSumTS} presents the coefficients used in \ac{ts} mode, where the rice parameter depends only on the top and left neighbor coefficients set $S_2$. When the remainder to encode is strictly below the threshold $\beta$, the \ac{trp} binarization is preformed with $p=cRiceParam$. Otherwise, the limited \ac{egk} coding is applied.

Algorithm~\ref{limitedEGk} shows that the maximum length of the prefix $maxPrefixLen$ depends on the range of the transform coefficients $2^{log2TrRange}$ and $BinReduc$. To differentiate between the two binarizations at the decoder side, $BinReduc$ is added to the prefix length when $Limited\_EGk$ is used. Then, a classical \ac{egk} binarization starts. However, if the computed prefix length $prefixLen$ is equal to the maximal prefix length $maxPrefixLen$, the suffix length $suffixLen$ is set to $log2TrRange$. Both codes are composed of a variable-length $prefix$ and if exists, a fixed-length $suffix$. The prefix is coded using a \ac{unary} or \ac{tucode} code representation which implies that changing any bin will violate the decoder standard or change the bitrate. On the other hand, the suffix might be encrypted in format compliance and constant bitrate only when the $LocSumAbs$ does not change the $cRiceParam$ value of the neighbor coefficients.}

\begin{algorithm}
\caption{$abs\_remainder$ Binarization}
	\renewcommand{\algorithmicrequire}{\textbf{Input:}}
	\renewcommand{\algorithmicensure}{\textbf{Output:}}
	\renewcommand{\algorithmiccomment}{\textbf{//}}
	\begin{algorithmic}
	\addcomment{
		\REQUIRE 
			$abs\_remainder$ is the unsigned integer to binarize\\
			$cRiceParam$ is the rice parameter\\
			$log2TrRange$ is the $log2$ of the \ac{tc} range\\
			$BinReduc \gets 5$ is the value used to determine the threshold between TRp and $Limited\_EGk$
		\STATE
		\STATE $\beta \gets BinReduc \; 2^{cRiceParam}$
		\IF {$ abs\_remainder < \beta$}
			\STATE	TRp binarization with $p \gets cRiceParam$
		\ELSE
			\STATE $Limited\_EGk(abs\_remainder,cRiceParam,$\\
			\hspace*{3 mm} $BinReduc, log2TrRange)$
		\ENDIF
	}
    \end{algorithmic}
\label{TcBinariation}
\end{algorithm}

\begin{algorithm}
\caption{$Limited\_EGk(abs\_remainder,cRiceParam, $\\$BinReduc, log2TrRange)$}
	\renewcommand{\algorithmicrequire}{\textbf{Input:}}
	\renewcommand{\algorithmicensure}{\textbf{Output:}}
	\renewcommand{\algorithmiccomment}{\textbf{//}}
	\begin{algorithmic}
    \addcomment{
		\REQUIRE 
			$abs\_remainder$, $cRiceParam$, $log2TrRange$, $BinReduc$.
		\STATE
		\STATE $maxPrefixLen \gets 32 - BinReduc - log2TrRange$
		\STATE $codeValue \gets \left \lfloor \frac{abs\_remainder}{2^{cRiceParam}} \right \rfloor - BinReduc$
        \STATE $prefixLen \gets 0$
        \WHILE { $prefixLen < maxPrefixLen$ \\\AND $codeValue > 2^{prefixLen+1} - 2  $  }
        \STATE $prefixLen \gets prefixLen + 1$
        \ENDWHILE
        \IF {$prefixLen = maxPrefixLen$}
        \STATE $suffixLen \gets log2TrRange$
        \ELSE
        \STATE $suffixLen \gets prefixLen + cRiceParam + 1$
        \ENDIF
        \STATE $totalPrefixLen  \gets prefixLen + BinReduc$
        \STATE $bitMask            \gets 2^{cRiceParam}$
        \STATE $prefix             \gets 2^{totalPrefixLen} - 1$
        \STATE $suffix             \gets codeValue -  2^{prefixLen}  - 1$
        \STATE $suffix \gets suffix \; 2^{cRiceParam} $\\
        \hspace*{3 mm} $ + abs\_remainder \mod bitMask$\\
        \COMMENT  {where $a\mod n$ gives the remainder of the euclidean divison of $a$ by $n$ }
}
    \end{algorithmic}
\label{limitedEGk}
\end{algorithm}





\section{Proposed \ac{vvc} selective encryption}
\label{sec:solution}
This section presents a new selective encryption scheme for \ac{vvc} standard. The proposed selective encryption fulfills two important features: standard format-compliant encryption (i.e. the bitstream must be decodable by any \ac{vvc} decoder) and constant bitrate encryption (i.e. preserve the \ac{vvc} compression efficiency).

The encryption is performed at the \ac{cabac} level of the encoder. Fig.~\ref{CACABEngine} depicts in green the position of the selective encryption in the \ac{cabac} engine. The encryption is performed after the binarization process, and only a set of selected syntax elements, listed in Table~\ref{syntaxencrypt} are ciphered. 
The encryption involves syntax elements from different coding tools including transform block, intra and inter predictions, and in-loop filters. This ensures the  encryption of both intra (\ac{I}) and inter (\ac{P} and \ac{B}) coded slices included in the \ac{vvc} video sequence.     

\begin{table}[ht]

\renewcommand{\arraystretch}{1.3} 
\caption{Encrypted syntax elements in the proposed \ac{vvc} selective encryption solution, all these syntax elements are bypass coded.} 
\label{syntaxencrypt} 
\centering 
\begin{tabular}{|l|l|c|}
\hline 
\bfseries {}  \bfseries Coding block & \bfseries Syntax elements & \bfseries Binarization \\
\hline 
\hline  Transform coefficients & {abs{\_}remainder, dec{\_}abs{\_}level} &{\ac{trp},\ac{egk}}  \\
\cline{2-3} (TCs) & {abs{\_}remainderTS} &{\ac{trp},\ac{egk}}  \\
\cline{2-3}  & {coeff{\_}sign{\_}flag} &{\ac{fl}}  \\
\cline{2-3}  & {coeff{\_}sign{\_}flagTS} &{\ac{fl}}  \\
\hline  \Acf{mv} & {abs{\_}mvd{\_}minus2} &{\ac{egk}}  \\
\cline{2-3}  & {mvd{\_}sign{\_}flag} &{\ac{fl}}  \\
\hline  ALF Filter & {alf{\_}luma{\_}fixed{\_}filter{\_}idx} &{\ac{tb}}  \\
\hline  Inter Prediction & {mmvd{\_}direction{\_}idx} &{\ac{fl}}  \\
\cline{2-3} & {merge{\_}triangle{\_}split{\_}dir} &{\ac{fl}}  \\
\hline  \ac{sao} Filter & sao\_offset\_sign & \ac{fl} \\
\cline{2-3}  & sao\_band\_position & \ac{fl} \\
\cline{2-3}  & sao\_eo\_class & \ac{fl} \\
\hline  Intra Prediction & intra\_chroma\_pred\_cand & \ac{fl} \\
\hline
\end{tabular} 
\end{table}

The syntax elements, listed in Table~\ref{syntaxencrypt}, have been selected based on following two criteria:
\begin{itemize}
    \item The syntax element is bypassed: this restriction preserves the \ac{vvc} coding efficiency. 
    \item Changing any bin will not change how the binstring is read by the decoder: this restriction ensures format-compliant encryption by excluding most of flags and syntax elements binarized by variable length codes.
\end{itemize}

The encryption of the most syntax elements listed in Table~\ref{syntaxencrypt} is straightforward except the \acp{tc} that requires a specific processing to search for the encryptable bins. In the next section, we describe the encryption of the \acp{tc} since it is the most challenging syntax element to encrypt. The  coding of the \acp{tc} introduces dependencies that need to be carefully addressed to perform format-compliant and constant bitrate encryption.      


\subsection{\sc{\acl{tc} encryption}}
\label{sec:solutionTC}
This section presents how the \acp{tc} are encrypted. As explained in Section~\ref{subsec:bin}, the binarization of the \acp{tc} and especially the length of the suffix depends on the previously encoded \acp{tc}. Therefore, encryption that changes the value of the coefficients may introduce bitrate increase. Indeed, the binarization depends on a rice parameter $cRiceParam\in \{ 0, 1, 2, 3\}$ derived from previous \acp{tc}. This rice parameter defines the fixed length of the suffix and therefore it corresponds to the size of the encryptable bins. After an analysis of the binarization algorithm, multiple conditions ensuring constant bitrate have emerged and are presented below.

First, it is important to note that the coefficients are binarized in two different ways depending on whether they are processed by pass 2-1 or 2-2, as presented in Fig.~\ref{TCsCoding}.
\begin{itemize}
\item \textit{The encryption \textbf{must not} change the parity of the coefficient:} changing the parity will result in changing the \textit{state} of the \ac{cabac} context. \addcomment{The state is updated using the previous state value and the parity of the current coefficient.}
\item \textit{The encryption \textbf{must not} change the rice parameter:} this will affect the bitrate.
\item \textit{The encryption \textbf{must not} change the $V$ value for coefficients processed by pass 2-2:} changing the value of this parameter can result in changing the parity, and thus the \ac{cabac} context.
\end{itemize}
\begin{figure}[t]
\vspace{-1em}
\centering
\subfloat[\label{LocSumAbs}]{\resizebox{0.40\linewidth}{!}{\begin{tikzpicture}[thick]

\tikzset{square matrix/.style={
    matrix of nodes,
    column sep=-\pgflinewidth, row sep=-\pgflinewidth,
    nodes={draw,
      minimum height=#1,
      anchor=center,
      text width=#1,
      align=center,
      inner sep=0pt
    },
  },
  square matrix/.default= 1.2cm
}

\matrix[square matrix]
{
\  				             & \                              & |[fill=green]| $X_c$ $Y_c-2$ 		   & \  \\
\                            & |[fill=green]| $X_c-1$ $Y_c-1$ & |[fill=green]| $X_c$ $Y_c-1$ 		   & \  \\
|[fill=green]| $X_c-2$ $Y_c$ & |[fill=green]| $X_c-1$ $Y_c$   & |[fill=yellow]|$\ \ \ X_c\ \ \ $ $Y_c$ & \  \\
\ 				             & \                              & \                 			           & \  \\
};
\node (0) at (-2.5,2.5) {$0$};
\node (O) at (-2.4,2.4) {};
\node (X) at (2.9,2.4) {$X$};
\node (Y) at (-2.4,-2.9) {$Y$};
\draw [->] {(O)  -- (X)};
\draw [->] {(O)  -- (Y)};
\end{tikzpicture}}}\hspace{5mm}
\subfloat[\label{LocSumAbsTS}]{\resizebox{0.40\linewidth}{!}{\begin{tikzpicture}[thick]

\tikzset{square matrix/.style={
    matrix of nodes,
    column sep=-\pgflinewidth, row sep=-\pgflinewidth,
    nodes={draw,
      minimum height=#1,
      anchor=center,
      text width=#1,
      align=center,
      inner sep=0pt
    },
  },
  square matrix/.default= 1.2cm
}

\matrix[square matrix]
{
\  				             & \                              & |[fill=red]| $X_c+1$ $Y_c-1$ 		   & \  \\
\                            & |[fill=yellow]| $X_c$ $Y_c$ & |[fill=green]| $X_c+1$ $Y_c$ 		   & \  \\
|[fill=red]| $X_c-1$ $Y_c+1$ & |[fill=green]| $X_c$ $Y_c+1$   & \  & \  \\
\ 				             & \                              & \                 			           & \  \\
};
\node (0) at (-2.5,2.5) {$0$};
\node (O) at (-2.4,2.4) {};
\node (X) at (2.9,2.4) {$X$};
\node (Y) at (-2.4,-2.9) {$Y$};
\draw [->] {(O)  -- (X)};
\draw [->] {(O)  -- (Y)};
\end{tikzpicture}}}
\caption{The current coefficient (in yellow) is used to compute the local absolute sum of the coefficients highlighted in green for (\ref{LocSumAbs}) \ac{tc} mode and (\ref{LocSumAbsTS})  \ac{ts} mode. Coefficients highlighted in red are used along the current coefficient in the prediction of the parity of the coefficients in green. $\bar{S}_1$ and $\bar{S}_2$ are two sets of green coefficients in (\ref{LocSumAbs}) and (\ref{LocSumAbsTS}), respectively.}
\label{LocSumAbsall}
\end{figure}
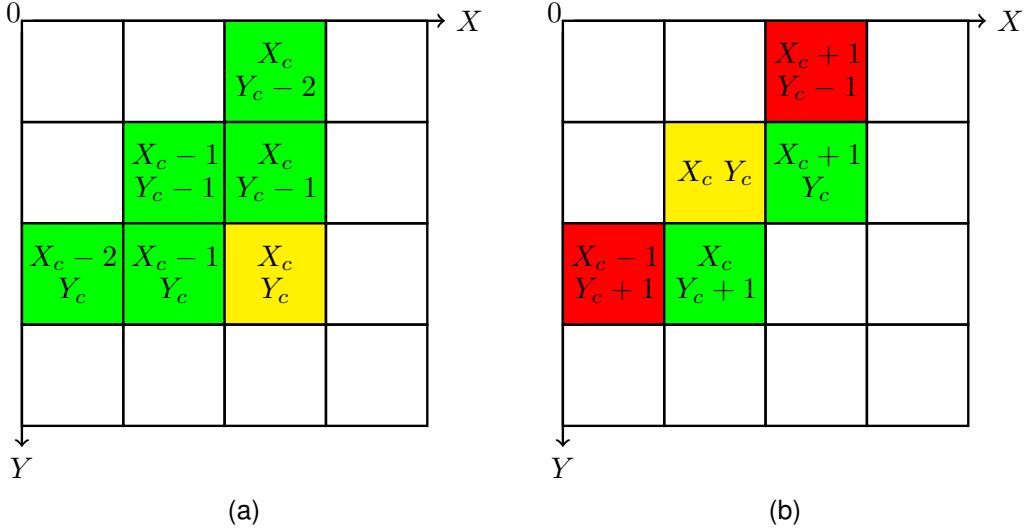
Considering those conditions, Algorithm~\ref{alg:TCencryptable} is proposed to identify the bins within the \acp{tc} that can be encrypted in constant bitrate and format compliance. The rice parameter of each \ac{tc} is derived from a saturated absolute sum of the local neighborhood of the current \ac{tc}. Fig.~\ref{LocSumAbs} depicts in green the set $\bar{S}_1$ of affected \acp{tc} if the current \ac{tc} in yellow is modified by encryption. Therefore, for each affected coefficient, Algorithm~\ref{alg:TCchecksum} checks whether the changes in the local absolute sum will affect the context, the rice parameter and the $V$ value. To make sure that the parity is not changed, the encryption excludes the \addcomment{\ac{lsb}} of the suffix and will perform encryption only when the rice parameter is greater than 1 ($cRiceParam \in \{ 2, 3\}$).

The encryption of the coefficients in \ac{ts} mode is similar. The main difference lies in how the rice parameter is derived. Fig.~\ref{LocSumAbsTS} shows the set $\bar{S}_2$ of affected coefficients in green when the current coefficient (in yellow) is modified by encryption. The current coefficient is binarized using a prediction based on the top and left coefficients. Therefore, the ciphered value of the current coefficient must remain lower or equal than the coefficient depicted in red in Fig.~\ref{LocSumAbsTS} according to the used prediction scheme.

\begin{algorithm}
\caption{$nbEncryptable = isEncryptable($ \\
$X_c, Y_c, CoeffArr, nbEncryptable, bypass, V)$}

\renewcommand{\algorithmicrequire}{\textbf{Input:}}
\renewcommand{\algorithmicensure}{\textbf{Output:}}
\renewcommand{\algorithmiccomment}{//}
\begin{algorithmic}[ht]
\REQUIRE
    $(X_c, Y_c)$: the coordinate of the current pixel,\\
    $CoeffArr[][]$: the array of coefficient value,\\
    $nbEncryptable$: the number of encryptable bits to test,\\
    $bypass$: \TRUE \ if the current coefficient is bypass,\\
    $V$: of the current pixel, set to $0$ if $bypass$ is \FALSE.\\
\ENSURE the number of encryptable bits.
\STATE
\STATE $absLevel \gets | CoeffArr[X_c][Y_c] |$
\IF {$absLevel \neq 0$ \AND $cRiceParam > 1$}
    \STATE $absCMin, remMin \gets computeMin(absLevel, $\\
    \STATE \hspace*{6 mm} $nbEncryptable, bypass, V)$\\
    \STATE $absCMax, remMax \gets computeMax(absLevel, $\\
    \STATE \hspace*{6 mm} $nbEncryptable, bypass, V)$\\
    \STATE $encryptable \gets$ \NOT ($bypass$ \\
        \hspace*{6 mm}  \AND $V \in [remMin, remMax]$)
    \FOR {$p \in \bar S_1$}
        \STATE $encryptable \gets encryptable$ \AND\\
                \hspace*{3 mm} $checkSumChange(X_{p}, Y_{p},$\\
                \hspace*{3 mm} $absLevel, absCMin, absCMax)$
    \ENDFOR\\
    \IF {\NOT $encryptable$ \AND $nbEncryptable > 1$}
        \RETURN $isEncryptable(X_c, Y_c, CoeffArr,$ \\
        \STATE \hspace*{3 mm} $nbEncryptable-1, bypass, V)$
    \ELSIF{$encryptable$}
        \RETURN $nbEncryptable$
    \ELSE
        \RETURN $0$
    \ENDIF
\ELSE
    \RETURN $0$
\ENDIF
\end{algorithmic}

\label{alg:TCencryptable}
\end{algorithm}

\begin{algorithm}
\footnotesize
\caption{$Encryptable = checkSumChange($\\
$X_p,Y_p, absLevel, absCMin, absCMax)$;}

\renewcommand{\algorithmicrequire}{\textbf{Input:}}
\renewcommand{\algorithmicensure}{\textbf{Output:}}
\renewcommand{\algorithmiccomment}{\textbf{//}}
\begin{algorithmic}
\REQUIRE
    $(X_p, Y_p)$: the coordinate of the tested coefficient ,\\
    $absLevel$: the absolute value of the coefficient $(X_c, Y_c)$ before encryption,\\
    $absCMin$: the minimal possible encrypted value of the coefficient $(X_c, Y_c)$,\\
    $absCMax$: the maximal possible encrypted value of the coefficient $(X_c, Y_c)$.
\ENSURE \TRUE\ if ciphered value does not affect the context, the rice parameter and the $C_0$, \FALSE\ otherwise.
    \STATE \COMMENT {\color{red} Step 1}
    \STATE $AbsSumP1 \gets \sum_{i \in S_1} \min(4 + |C_i|\mod 2, |C_i|)$\\
    \STATE $numPos \gets \sum_{i \in S_1} E(C_i)$\\
    \COMMENT  {where $E(x)$ returns 1 if $x \neq 0$, 0 otherwise }
    \STATE $AbsSumP1min \gets AbsSumP1 $\\
    \hspace*{3 mm} $- \min(4 + (absLevel \mod 2), absLevel) $\\
    \hspace*{3 mm} $+ \min(4 + (absCMin \mod 2), absCMin)$
    \STATE $NoCtxChange \gets \left \lfloor \frac{AbsSumP1+1}{2}\right \rfloor \geq 3$ \\
    \hspace*{3 mm} \AND $AbsSumP1 - numPos \geq 4$\\
    \hspace*{3 mm} \AND $\left \lfloor \frac{AbsSumP1Min+1}{2}\right \rfloor  \geq 3$\\
    \hspace*{3 mm} \AND $AbsSumP1Min - numPos \geq 4$
    \STATE \COMMENT {\color{red} Step 2}
    \STATE $AbsSumP21 \gets \sum_{i \in S_1} |C_i|$
    \STATE $AbsSumP21Min \gets AbsSumP21-absLevel$\\
    \hspace*{3 mm}$+absCMin$
    \STATE $AbsSumP21Max \gets AbsSumP21-absLevel$\\
    \hspace*{3 mm}$+absCMax$
    \STATE $TrAbsSumP21 \gets [AbsSumP21 - 20]^{31}_{0}$
    \STATE $TrAbsSumP21Min \gets [AbsSumP21Min - 20]^{31}_{0}$
    \STATE $TrAbsSumP21Max \gets [AbsSumP21Max - 20]^{31}_{0}$
    \STATE $RiceParP21 \gets riceArr[TrAbsSumP21]$
    \STATE
    \STATE $TrAbsSumP22 \gets [AbsSumP21]^{31}_{0}$
    \STATE $TrAbsSumP22Min \gets [AbsSumP21Min]^{31}_{0}$
    \STATE $TrAbsSumP22Max \gets [AbsSumP21Max]^{31}_{0}$    
    \STATE $RiceParP22 \gets riceArr[TrAbsSumP22]$
    \STATE \COMMENT {\color{red} Step 3}
    \STATE $NoRiceParChange  \gets \TRUE$
    \IF {$TrAbsSumP21Min \notin I_R[RiceParP21]$\\
         \OR $TrAbsSumP21Max \notin I_R[RiceParP21]$\\
         \OR $TrAbsSumP22Min \notin I_R[RiceParP22]$\\
         \OR $TrAbsSumP22Max \notin I_R[RiceParP22]$}
    \STATE  \hspace*{3 mm} $NoRiceParChange  \gets \FALSE$
    \ENDIF
    \STATE \COMMENT {\color{red} Step 4}
    \STATE  $NoVChange  \gets \TRUE$
    \FOR{$i\gets 0$ \TO $2$}
    \STATE $currV \gets VArr[i][TrAbsSumP22]$
    \IF {$TrAbsSumP22Min \notin I_P[i][currV]$\\
         \OR $TrAbsSumP22Max \notin I_P[i][currV]$}
    \STATE \hspace*{3 mm} $NoVChange \gets \FALSE$
    \ENDIF
    \ENDFOR
    \STATE \COMMENT {\color{red} Step 5}
    \RETURN $NoCtxChange$ \AND $NoRiceParChange$ \AND $NoVChange$

\end{algorithmic}

\label{alg:TCchecksum}
\end{algorithm}



Algorithms~\ref{alg:TCencryptable} and~\ref{alg:TCchecksum} are used to check that the encrypted bins of the current binarized coefficient are not affecting how the neighbor coefficients will be encoded. This enables defining the bins that can be encrypted in format-compliant and constant bitrate. The proposed solution is carried out as follows: 
\begin{itemize}
    \item Algorithm~\ref{alg:TCencryptable} checks at the binarization process whether the \ac{tc} can be encrypted or not.  The encryption is possible only when the absolute value of the coefficient is different from 0 ($absLevel \neq 0$) and the value of the derived rice parameter is above 1 ($cRiceParam > 1$).
    \item The algorithm then computes the minimum and maximum values of the encrypted remainder ($remMin, remMax$) of the current coefficient. The minimum ($absCMin$) and maximum ($absCMax$) absolute values of the coefficient are derived from their respective remainders.
    \item Algorithm~\ref{alg:TCchecksum} checks for all coefficients in $\bar S_1$ depicted in green in Fig.~\ref{LocSumAbs} (when they exist) whether ciphering the current coefficient $(X_c, Y_c)$ will affect its neighbor coefficients $(X_p, Y_p)$, the rice parameter or the $V$ value. This operation is performed in five steps as follows: 

    \begin{enumerate}
        \item The algorithm computes a saturated absolute sum of the tested coefficient of coordinates $(X_p, Y_p)$ ($AbsSumP1 = \sum_{i \in S_1} min(4 + |C_i| \mod 2, |C_i|)$) which is used for the context computation, with $S_1$ the set of neighbor coefficients of $(X_p, Y_p)$. This operation is performed at the first pass (P1) to check the \ac{cabac} context change and set the $NoCtxChange$ flag to \textbf{true} if the changes on the $AbsSumP1$ will not affect the context.
        \item The local absolute sum \addcomment{$TrAbsSumP21$} is then computed by~\eqref{eq:localabssum} for the coefficient of coordinates $(X_p, Y_p)$. The minimum possible value \addcomment{$TrAbsSumP21Min$} and the maximum value \addcomment{$TrAbsSumP21Max$} are also computed by~\eqref{eq:localabssum} with $BaseLvl$ equals to 4. 
        \item Then, the algorithm checks whether the rice parameter will be affected with the different computed sums in step 2 and sets a flag $NoRiceParChange$ to \textbf{true} if the rice parameter remains unchanged with all possible tested conditions. \addcomment{$I_R$ is a \ac{lut} containing, for each rice value, the interval in which the the local absolute sum does not change the rice parameter}.
        \item The parameter $V$ is computed only in pass 2-2. 
        At this fourth step, the algorithm checks for the processed coefficients if the parameter $V$ remains unchanged to set the flag $NoVChange$ to \textbf{true}. 
        \addcomment{Similar to $I_R$, $I_P$ returns, depending of the state and $V$ values, the interval in which the local absolute sum does not change the $V$ value}.
        \item Finally, Algorithm~\ref{alg:TCchecksum} returns \textbf{true} when $NoCtxChange$, $NoRiceParChange$ and $NoVChange$ are all equal to \textbf{true}.
    \end{enumerate}

\end{itemize}

The decoder performs inverse operations performed by the encoder for deciphering. The decoder first decodes the \acp{tc} and then it searches for the encryptable coefficients using Algorithms~\ref{alg:TCencryptable} and~\ref{alg:TCchecksum}. Finally, the deciphering will process only the identified encrypted bins. 


\subsection{\sc{Encryption Method and Synchronisation}}
The syntax elements to cipher are now defined. To cipher the syntax elements of a variable  length, a stream cipher is more suited for this application. As the minimum error propagation is one of the most desirable properties in video encryption, we use the \ac{aes} algorithm in \ac{ctr} mode as a \ac{prng} to encrypt the identified syntax elements. It is important to note that \ac{ctr} counter value should not be reused, which is adopted in our solution~\cite{lipmaa2000comments}. Meanwhile, other stream ciphers such as Rabbit~\cite{boesgaard_rabbit_2008}, \ac{lwcb} stream cipher~\cite{gautier:hal-02184571}, HC-128~\cite{wu_stream_2008} or even block ciphers like \ac{aes} in \ac{cfb} mode, can be used as well.
A stream cipher produces a cipher text $C$ using a \ac{xor} operation between the plain text $P$ and the output steam $X_g$ produced by a \ac{prng},
\begin{equation}
    C(P)= P \oplus X_g.
\end{equation}


To revert the encryption, a \ac{xor} between the cipher text and the same \ac{prng} output is performed. Thus, a perfect synchronization between the encoder and the decoder is required.
Most of the syntax elements are systematically ciphered and do not dependant on the position or the context. However, the syntax elements associated to the \acp{tc} are ciphered only if they meet conditions previously described in Section~\ref{sec:solutionTC}. One of this conditions relies on the neighbor coefficients, implying that the last decoded coefficient needs to be deciphered first. To allow this behavior, for each significant coefficient, encryptable or not, the \ac{prng} generates a sample equal to the size of the rice parameter, e.g. the maximum encryptable size. The unused samples are discarded to keep the encoder and the decoder perfectly synchronized.
In \ac{ctr} mode, one bit flipping caused by transmission errors will only affect one bit during the deciphering process which minimizes the error propagation.


\section{Results and Discussions}
\label{sec:results}
In this section, we first present the experimental setup, followed by an assessment of the video degradation introduced by the selective encryption, then a security analysis will be presented, and finally a complexity evaluation is provided.
\subsection{\sc{Experimental Setup}}
The experiments are carried-out under the \acp{ctc} of the \ac{vvc} standard. The \acp{ctc} define several test video sequences of different resolutions, and five \acp{qp} are used $\acs{qp} \in \{17, \, 22,  \, 27, \, 32, \, 37 \}$. 
The proposed encryption solution is implemented in the \ac{vtm}~\cite{VtmGitlab} version 6.0. \Ac{vtm} is the reference software implementation of both encoder and decoder of the \ac{vvc} standard. \addcomment{The coding configuration without encryption is referred to as the Anchor}. The video sequences are encoded with encryption in Random Access (RA) coding configuration. This latter is the common coding configuration used in broadcast and \ac{ott} applications with an Intra period of 32 frames.     The complexity measurements are performed on a desktop computer equipped with an Intel i7-7700 processor running at 3.60 GHz on Ubuntu 18.04 OS.  

\begin{table*}[htb]
\tiny
\centering
\caption{\ac{psnr} performance of the proposed selective encryption for all video sequences at five \ac{qp}s. Anchor and ciphered configurations correspond to the video decoded without encryption and with selective encryption, respectively.}

\begin{tabular}{|c|l||c|c||c|c||c|c||c|c||c|c|}
\hline
\multicolumn{2}{|c||}{}& \multicolumn{10}{c|}{$PSNR$ Scores (dB)} \\ \cline{3-12}
\multicolumn{2}{|c||}{} & \multicolumn{2}{c||}{$QP\ 17$} & \multicolumn{2}{c||}{$QP\ 22$} & \multicolumn{2}{c||}{$QP\ 27$} & \multicolumn{2}{c||}{$QP\ 32$} & \multicolumn{2}{c|}{$QP\ 37$} \\ \cline{3-12}
\multicolumn{2}{|c||}{} & Anchor & Ciphered & Anchor & Ciphered & Anchor & Ciphered & Anchor & Ciphered & Anchor & Ciphered\\ \hline
\hline
\multirow{3}{*}{A1} & Campfire & 44.02 & 4.58 & 39.78 & 4.69 & 37.64 & 4.60 & 36.54 & 4.38 & 35.18 & 4.37\\ 
\cline{2-12}
 & FoodMarket4 & 46.21 & 10.99 & 44.29 & 10.79 & 42.96 & 10.62 & 41.13 & 9.76 & 38.84 & 10.83\\ 
\cline{2-12}
 & Tango2 & 42.37 & 8.66 & 40.40 & 9.05 & 39.73 & 10.40 & 38.88 & 8.16 & 37.58 & 8.68\\ 
\hline
\hline
\multirow{3}{*}{A2} & CatRobot1 & 42.84 & 9.13 & 40.48 & 9.55 & 39.56 & 9.73 & 38.37 & 8.72 & 36.70 & 9.08\\ 
\cline{2-12}
 & DaylightRoad2 & 41.76 & 8.93 & 38.25 & 11.16 & 37.33 & 9.34 & 36.44 & 9.74 & 35.12 & 10.00\\ 
\cline{2-12}
 & ParkRunning3 & 47.51 & 11.24 & 43.83 & 11.52 & 39.85 & 10.56 & 36.58 & 12.00 & 33.60 & 11.10\\ 
\hline
\hline
\multirow{5}{*}{B} & BasketballDrive & 42.08 & 13.17 & 39.55 & 12.95 & 37.98 & 11.90 & 36.34 & 11.26 & 34.40 & 11.63\\ 
\cline{2-12}
 & BQTerrace & 42.76 & 10.22 & 37.66 & 10.23 & 35.56 & 10.28 & 34.29 & 10.15 & 32.78 & 9.82\\ 
\cline{2-12}
 & Cactus & 41.54 & 10.04 & 38.74 & 10.23 & 37.24 & 9.48 & 35.59 & 9.12 & 33.51 & 8.99\\ 
\cline{2-12}
 & MarketPlace & 43.52 & 9.55 & 40.96 & 9.10 & 38.84 & 8.22 & 36.72 & 8.61 & 34.44 & 8.67\\ 
\cline{2-12}
 & RitualDance & 47.02 & 9.43 & 44.79 & 10.44 & 41.71 & 10.11 & 38.71 & 9.28 & 35.76 & 9.56\\ 
\hline
\hline
\multirow{4}{*}{C} & BasketballDrill & 44.19 & 13.40 & 41.69 & 12.83 & 38.57 & 12.88 & 35.71 & 11.94 & 33.09 & 11.22\\ 
\cline{2-12}
 & BQMall & 42.70 & 11.94 & 40.73 & 11.12 & 38.34 & 10.79 & 35.81 & 11.16 & 33.14 & 10.65\\ 
\cline{2-12}
 & PartyScene & 42.13 & 12.02 & 39.05 & 11.78 & 35.73 & 11.34 & 32.76 & 11.21 & 29.96 & 11.33\\ 
\cline{2-12}
 & RaceHorsesC & 43.13 & 11.53 & 39.58 & 11.58 & 36.48 & 11.74 & 33.80 & 11.64 & 31.15 & 10.97\\ 
\hline
\hline
\multirow{4}{*}{D} & BasketballPass & 45.22 & 13.54 & 41.58 & 13.44 & 37.54 & 13.15 & 34.27 & 13.41 & 31.39 & 13.79\\ 
\cline{2-12}
 & BlowingBubbles & 41.75 & 11.20 & 38.81 & 11.69 & 35.60 & 11.06 & 32.69 & 11.44 & 29.87 & 11.57\\ 
\cline{2-12}
 & BQSquare & 42.10 & 8.89 & 38.69 & 9.37 & 35.41 & 8.78 & 32.70 & 8.93 & 30.22 & 9.62\\ 
\cline{2-12}
 & RaceHorses & 43.66 & 11.75 & 40.14 & 12.11 & 36.55 & 12.05 & 33.25 & 11.74 & 30.31 & 11.56\\ 
\hline
\hline
\multirow{3}{*}{E} & FourPeople & 44.92 & 9.91 & 43.38 & 9.72 & 41.81 & 8.73 & 39.82 & 8.93 & 37.30 & 8.70\\ 
\cline{2-12}
 & Johnny & 44.99 & 9.54 & 43.53 & 9.26 & 42.38 & 8.70 & 40.94 & 8.58 & 39.01 & 8.48\\ 
\cline{2-12}
 & KristenAndSara & 45.42 & 9.50 & 43.86 & 9.15 & 42.40 & 7.60 & 40.64 & 7.73 & 38.42 & 7.62\\ 
\hline
\hline
\multirow{4}{*}{F} & ArenaOfValor & 46.94 & 11.15 & 43.63 & 10.78 & 40.28 & 10.67 & 37.45 & 9.38 & 34.79 & 9.46\\ 
\cline{2-12}
 & BasketballDrillText & 44.31 & 12.52 & 41.73 & 12.01 & 38.52 & 12.10 & 35.60 & 11.24 & 32.89 & 11.21\\ 
\cline{2-12}
 & SlideEditing & 54.72 & 9.69 & 51.35 & 10.49 & 47.31 & 10.15 & 43.25 & 10.72 & 38.92 & 9.31\\ 
\cline{2-12}
 & SlideShow & 56.12 & 1.94 & 52.43 & 3.16 & 48.61 & 4.28 & 45.17 & 3.98 & 41.83 & 4.35\\ 
\hline
\hline
\multicolumn{1}{|c}{} & \multicolumn{1}{l||}{\textbf{Average}} & \textbf{44.77} & \textbf{10.17} & \textbf{41.88} & \textbf{10.32} & \textbf{39.38} & \textbf{9.97} & \textbf{37.05} & \textbf{9.74} & \textbf{34.62} & \textbf{9.71}\\\hline
\end{tabular}

\label{psnr_complete}

\end{table*}

\begin{table*}[htb]
\tiny
\centering
\caption{Average \ac{ssim} performance of the proposed encryption solution for all video classes at five \ac{qp}s}


\begin{tabular}{|c||c|c||c|c||c|c||c|c||c|c|}
\hline
\multicolumn{1}{|c||}{}& \multicolumn{10}{c|}{$SSIM$ Score} \\ \cline{2-11}
\multicolumn{1}{|c||}{} & \multicolumn{2}{c||}{$QP\ 17$} & \multicolumn{2}{c||}{$QP\ 22$} & \multicolumn{2}{c||}{$QP\ 27$} & \multicolumn{2}{c||}{$QP\ 32$} & \multicolumn{2}{c|}{$QP\ 37$} \\ \cline{2-11}
\multicolumn{1}{|c||}{} & Anchor & Ciphered & Anchor & Ciphered & Anchor & Ciphered & Anchor & Ciphered & Anchor & Ciphered\\ \hline
\hline
A1 & 1.00 & 0.27 & 1.00 & 0.22 & 1.00 & 0.23 & 0.99 & 0.21 & 0.99 & 0.22\\  
\hline

A2 & 1.00 & 0.23 & 1.00 & 0.23 & 1.00 & 0.21 & 0.99 & 0.21 & 0.99 & 0.21\\ 
\hline

B & 1.00 & 0.27 & 1.00 & 0.27 & 0.99 & 0.26 & 0.99 & 0.25 & 0.97 & 0.25\\ 
\hline

C & 1.00 & 0.23 & 0.99 & 0.23 & 0.98 & 0.23 & 0.97 & 0.23 & 0.94 & 0.24\\ 
\hline

D & 0.99 & 0.24 & 0.97 & 0.24 & 0.95 & 0.25 & 0.91 & 0.26 & 0.85 & 0.28\\ 
\hline

E & 1.00 & 0.39 & 1.00 & 0.40 & 0.99 & 0.37 & 0.99 & 0.37 & 0.99 & 0.39\\ 
\hline

F & 1.00 & 0.26 & 1.00 & 0.31 & 1.00 & 0.35 & 0.99 & 0.36 & 0.97 & 0.35\\ 
\hline
\hline
\multicolumn{1}{|c||}{\textbf{Average}} & \textbf{1.00} & \textbf{0.27} & \textbf{0.99} & \textbf{0.27} & \textbf{0.99} & \textbf{0.27} & \textbf{0.97} & \textbf{0.27} & \textbf{0.95} & \textbf{0.28}\\\hline
\end{tabular}

\label{ssim_complete}
\end{table*}

\begin{table*}[htb]
\tiny
\centering
\caption{Average \ac{vmaf} performance of the proposed encryption solution for all video classes at five \ac{qp}s}


	\begin{tabular}{|c||c|c||c|c||c|c||c|c||c|c|}
		\hline
		\multicolumn{1}{|c||}{}& \multicolumn{10}{c|}{$VMAF$ Score} \\ \cline{2-11}
		\multicolumn{1}{|c||}{} & \multicolumn{2}{c||}{$QP\ 17$} & \multicolumn{2}{c||}{$QP\ 22$} & \multicolumn{2}{c||}{$QP\ 27$} & \multicolumn{2}{c||}{$QP\ 32$} & \multicolumn{2}{c|}{$QP\ 37$} \\ \cline{2-11}
		\multicolumn{1}{|c||}{} & Anchor & Ciphered & Anchor & Ciphered & Anchor & Ciphered & Anchor & Ciphered & Anchor & Ciphered\\ \hline
		\hline
			A1 	& 99.47	& 25.53	& 99.03	& 28.84	& 96.70	& 30.23	& 92.31	& 29.84	& 84.86	& 28.19\\\hline
			A2 	& 99.88	& 23.79	& 99.22	& 23.88	& 97.20	& 25.15	& 92.49	& 25.27	& 85.02	& 25.22\\\hline
			B 	& 99.80	& 2.78	& 99.08	& 2.89	& 96.35	& 3.24	& 89.46	& 3.52	& 77.84	& 3.67\\\hline
			C 	& 99.88	& 7.16	& 99.59	& 7.72	& 96.91	& 7.78	& 89.84	& 7.63	& 77.55	& 7.88\\\hline
			D 	& 99.42	& 6.66	& 98.78	& 6.57	& 95.68	& 6.76	& 87.96	& 6.75	& 75.68	& 6.29\\\hline
			E 	& 97.37	& 0.16	& 96.54	& 0.51	& 95.00	& 0.62	& 92.04	& 0.71	& 86.52	& 1.27\\\hline
			F 	& 98.88	& 6.20	& 98.65	& 7.83	& 96.72	& 8.57	& 92.47	& 8.66	& 85.43	& 6.86\\\hline
		\hline
		\multicolumn{1}{|c||}{\textbf{Average}} & \textbf{99.30} & \textbf{9.32} & \textbf{98.76} & \textbf{10.10} & \textbf{96.37} & \textbf{10.64} & \textbf{90.73} & \textbf{10.66} & \textbf{81.27} & \textbf{10.25}\\\hline
	\end{tabular}

\label{vmaf_complete}
\vspace{-2em}
\end{table*}

\subsection{\sc{Video Quality and Encryption Space}}
\label{subsec:quality}
\subsubsection{\sc{Video Quality}} The distortion introduced by the proposed solution on the test video sequences is assessed in this section. Three full-reference objective image and video quality metrics are computed on the encrypted video sequences with respect to the original. \Ac{psnr} is used to evaluate the video quality based on the mean squared error computed over the frame pixels \cite{winkler2008evolution}. \addcomment{The \ac{psnr} is computed as a weighted sum of the \ac{psnr} scores of the three color components. 
}
\ac{ssim} explores the structural similarity between the original and the decoded frame. It is important to note that a \ac{ssim} value close to 1 refers to decoded frame of a similar quality as the original frame~\cite{wang2004image}. Finally, \ac{vmaf} is a video quality metric that predicts the perceived quality score of a video sequence~\cite{vmaf}, where a score of 100 indicates a good perceptual video quality and 0 refers to a very low perceived video quality.

\begin{figure*}
    \subfloat[\label{fig:anchor_qp17} $QP\ 17$, PSNR=42.90 dB]{\includegraphics[width=0.19\textwidth]{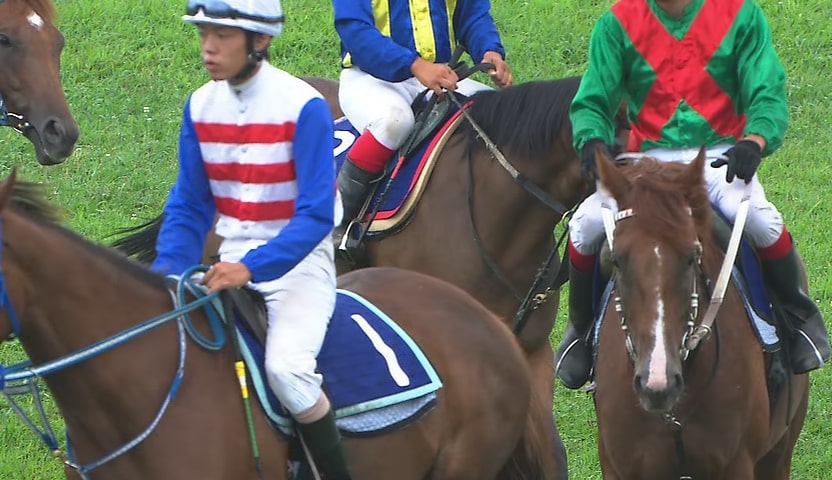}}     \hfill
    \subfloat[\label{fig:anchor_qp22} $QP\ 22$, PSNR=39.09 dB]{\includegraphics[width=0.19\textwidth]{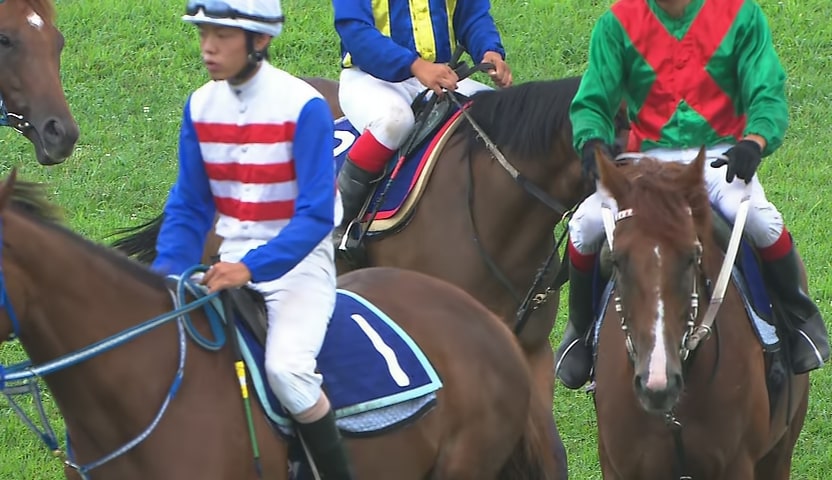}}     \hfill
    \subfloat[\label{fig:anchor_qp27} $QP\ 27$, PSNR=34.98 dB]{\includegraphics[width=0.19\textwidth]{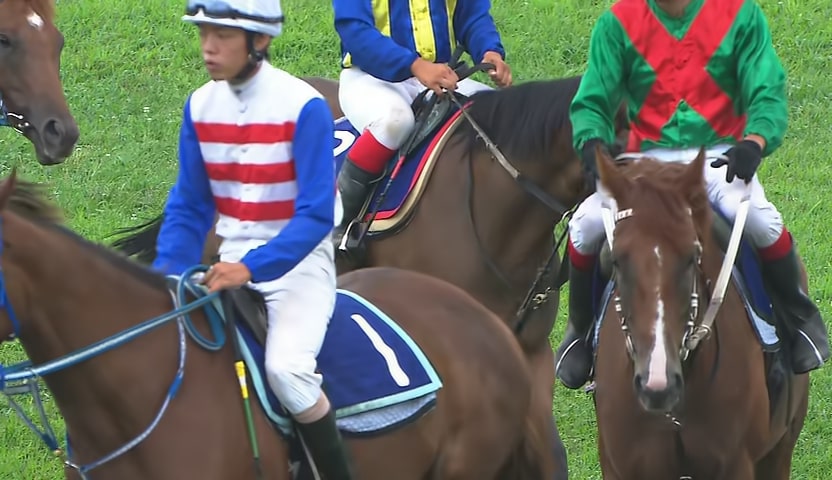}}     \hfill
    \subfloat[\label{fig:anchor_qp32} $QP\ 32$, PSNR=32.53 dB]{\includegraphics[width=0.19\textwidth]{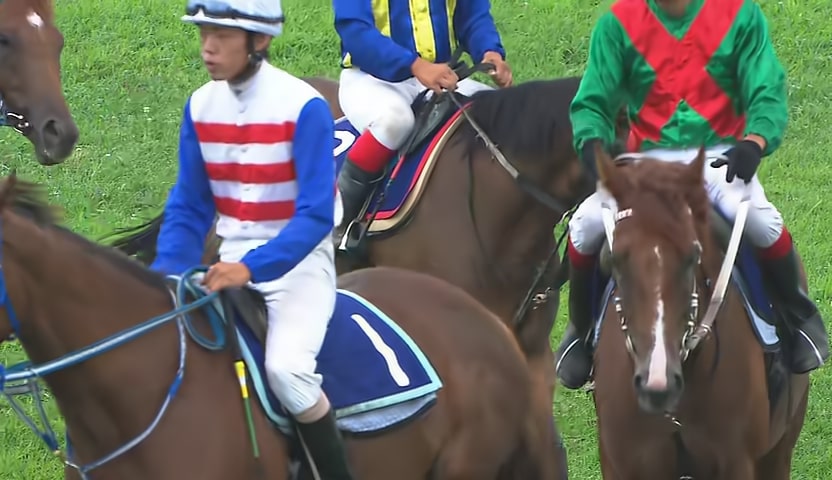}}     \hfill
    \subfloat[\label{fig:anchor_qp37} $QP\ 37$, PSNR=29.72 dB]{\includegraphics[width=0.19\textwidth]{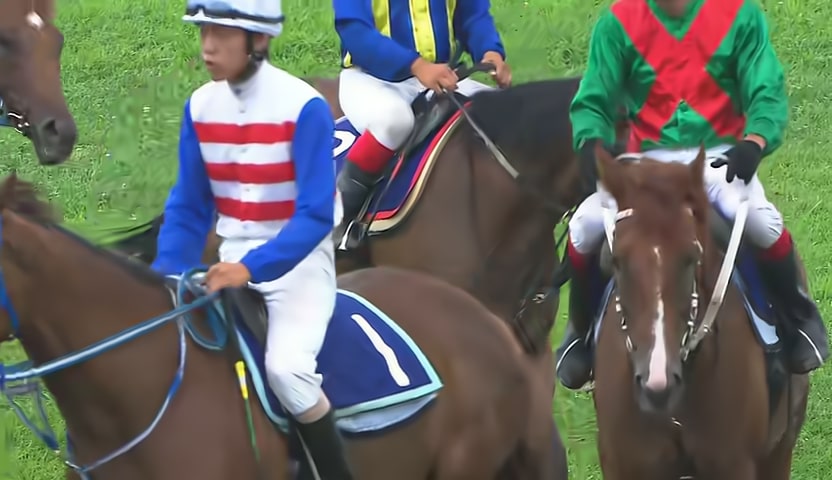}}     \hfill
    
    \subfloat[\label{fig:cipher_qp17} $QP\ 17$, PSNR=12.07 dB]{\includegraphics[width=0.19\textwidth]{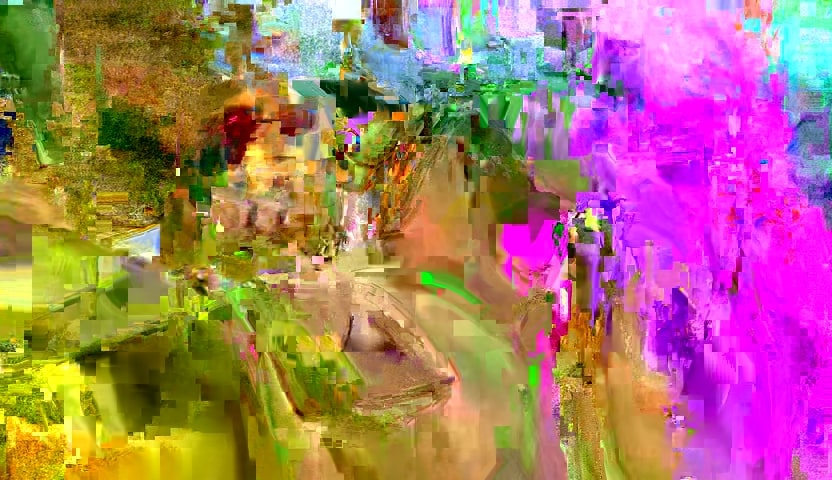}}     \hfill
    \subfloat[\label{fig:cipher_qp22} $QP\ 22$, PSNR=11.73 dB]{\includegraphics[width=0.19\textwidth]{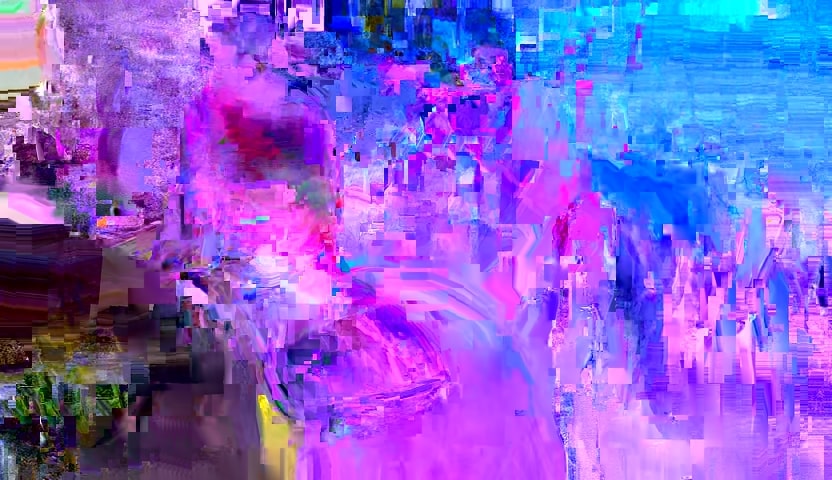}}     \hfill
    \subfloat[\label{fig:cipher_qp27} $QP\ 27$, PSNR=10.03 dB]{\includegraphics[width=0.19\textwidth]{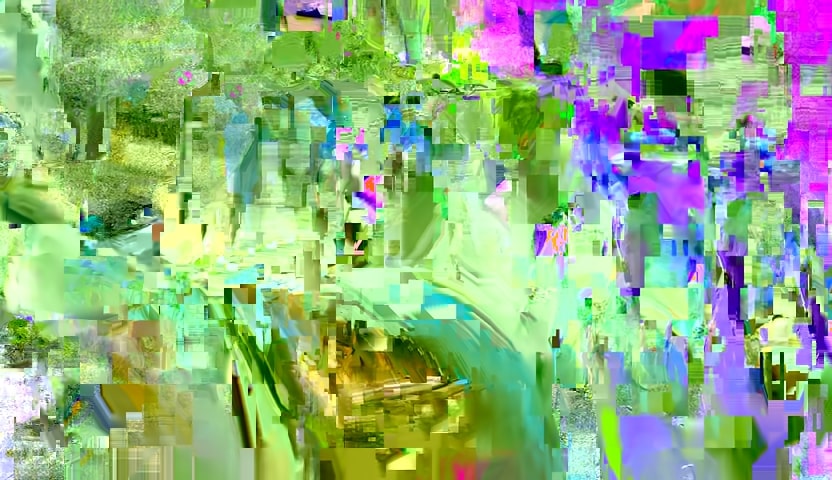}}     \hfill
    \subfloat[\label{fig:cipher_qp32} $QP\ 32$, PSNR=10.68 dB]{\includegraphics[width=0.19\textwidth]{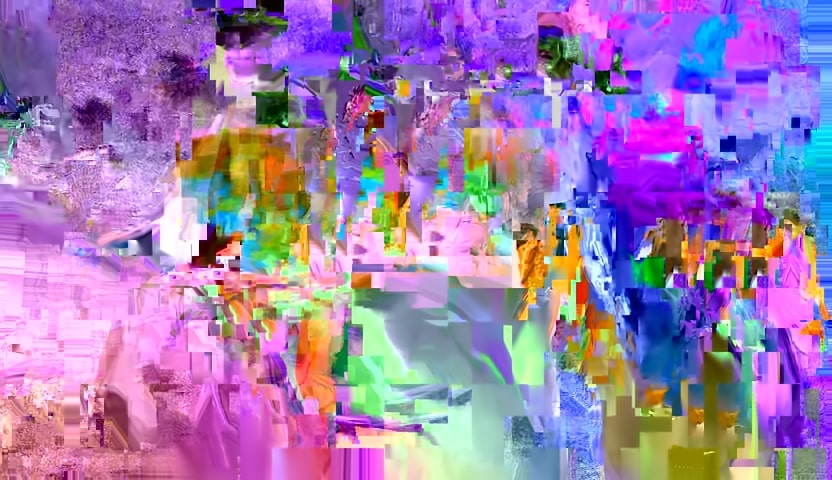}}     \hfill
    \subfloat[\label{fig:cipher_qp37} $QP\ 37$, PSNR=10.25 dB]{\includegraphics[width=0.19\textwidth]{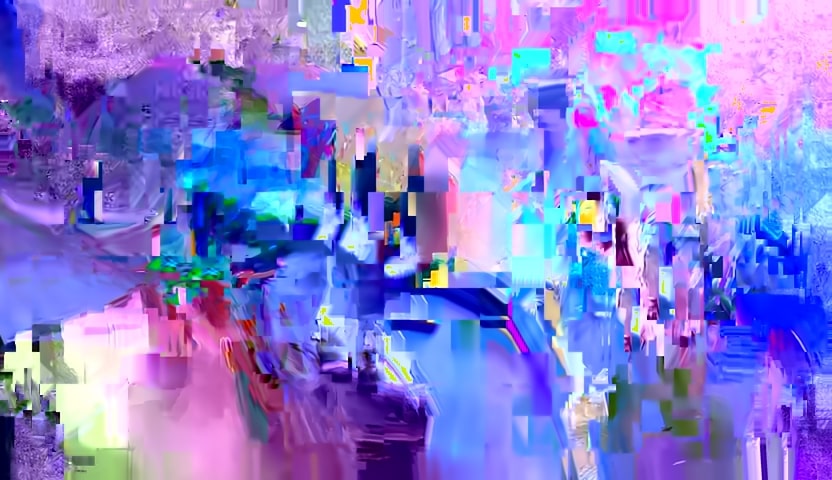}}     \hfill

    \caption{Visual illustration of Frame \#10 of {\it RaceHorsesC} video decoded without encryption (\ref{fig:anchor_qp17} - \ref{fig:anchor_qp37}) and with selective encryption (\ref{fig:cipher_qp17} - \ref{fig:cipher_qp37}) at five QPs}
    \label{fig:snapshot_qp_quality}
        \vspace{-1em}

\end{figure*}

Table~\ref{psnr_complete} presents the \ac{psnr} performance over all video sequences at the five considered \acp{qp}. We can notice that the \ac{psnr} drops at \ac{qp} 17 from 44.77 dB in average to around 10.17 dB. 
The same \ac{psnr} values of encrypted videos are reached on different \acp{qp} values. This indicates that the proposed solution significantly decreases the objective quality of the encrypted video. \addcomment{{\it Campfire} and {\it SlideShow} encrypted video sequences have a very low \ac{psnr} values. For \textit{Campfire}, it can be explained by its texture and complex shapes associated to high motion that increase the encryption space and thus improving the quality of the encryption. Concerning \textit{SlideShow}, shapes are less complex however the encryption is able to flip the colors causing noticeable quality degradation with lower \ac{psnr} scores than the average}.

Table~\ref{ssim_complete} presents the average \ac{ssim} scores for seven video classes at different \ac{qp}s. The proposed encryption solution enables to reduce the \ac{ssim} from around 1 to 0.25. The obtained \ac{ssim} value confirms that the proposed solution introduces a drastic distortion on the structure information within the encrypted video frame. \addcomment{We can notice that \ac{ssim} scores of class E video sequences are higher than the average scores. These video sequences have low motion and less texture compared to other sequences. This improves the coding efficiency and decreases the performance of the selective encryption since less syntax elements are encrypted.}

Finally, Table~\ref{vmaf_complete} presents the \ac{vmaf} scores which also emphasize the large degradation of the subjective video quality as a result of using the proposed encryption solution. 

Fig.~\ref{fig:snapshot_qp_quality} illustrates the frame \#10 of \textit{RaceHorcesC} video sequence decoded at five \ac{qp} values with and without encryption. The visual quality of decoded encrypted video is very low making difficult to recognize objects and colors in the video frame at all \ac{qp} values with \ac{psnr} scores around 11 dB.

\subsubsection{\sc{Encryption space}} \addcomment{The computational time of encryption mainly depends on the encryption space of any ciphering process.  However, the robustness and security level will be
enhanced by increasing the encryption space. In selective encryption, the somehow robust encryption algorithm and low computational overhead as outcome of the encryption is the target.} Table~\ref{ES_syntax} presents the encryption space of the proposed encryption solution as the percentage of encrypted bits by syntax element on the whole bitstream. The quality degradation of selective encryption is achieved by ciphering only 26.66\% and 15.42\% of the bitstream at high and low bitrates, respectively.  We can notice that the largest encryption space is enabled by the encryption of the \acp{tc} while the part of other syntax element less present in the bitstream remains negligible ($<$ 2\%).

\begin{table*}[htp]
\tiny
\centering

\begin{tabular}{|c||c|c|c|c|c|}

\hline
\multirow{3}{*}{Syntax Elements} & \multicolumn{5}{c|}{Encryption Space (\%)} \\\cline{2-6}
& $QP$ & $QP$ & $QP$ & $QP$ & $QP$ \\
& $17$ & $22$ & $27$ & $32$ & $37$  \\\hline
alf\_luma\_filter\_idx & 0.01 & 0.02 & 0.04 & 0.06 & 0.06 \\\hline
sao\_offset\_sign & 0.00 & 0.00 & 0.00 & 0.00 & 0.00 \\\hline
sao\_band\_position & 0.01 & 0.01 & 0.01 & 0.01 & 0.01 \\\hline
sao\_eo\_class & 0.01 & 0.02 & 0.01 & 0.01 & 0.00 \\\hline
mmvd\_direction\_idx & 0.48 & 0.55 & 0.56 & 0.59 & 0.59 \\\hline
merge\_triangle\_split\_dir & 0.04 & 0.08 & 0.13 & 0.17 & 0.19 \\\hline
mvd\_abs & 0.26 & 0.48 & 0.76 & 0.98 & 1.14 \\\hline
mvd\_sign & 0.29 & 0.51 & 0.75 & 0.92 & 1.00 \\\hline
\begin{tabular}{@{}c@{}}abs\_remainder \\ coeff\_sign\end{tabular} & 25.44 & 21.39 & 17.15 & 14.33 & 12.22 \\\hline
\begin{tabular}{@{}c@{}}abs\_remainderTS \\ coeff\_signTS\end{tabular} & 0.06 & 0.06 & 0.05 & 0.03 & 0.01 \\\hline
\begin{tabular}{@{}c@{}}intra\_chroma \\ \_pred\_candidate\end{tabular} & 0.07 & 0.10 & 0.14 & 0.16 & 0.18 \\\hline \hline
\textbf{Total} & \textbf{26.66} & \textbf{23.21} & \textbf{19.61} & \textbf{17.24} & \textbf{15.42} \\\hline
\end{tabular}

\label{ES_syntax}
\vspace{-1em}

\caption{Encryption space in percentage (\%) per syntax element at five \acp{qp}}

\end{table*}


\subsection{\sc{Security Analysis}}
In the previous section we only assess the visual degradation achieved by the proposed encryption. In this section we focus on the quality of the proposed encryption and its robustness against different types of attacks.

\subsubsection{\sc{\acf{eq} analysis}}
The algebraic summation of differences between pixels distributions of the original frame $H(P)$ and the encrypted fame $H(C)$ is called \acs{eq}. This latter is computed as follows~\cite{EQmetric}
 \begin{equation}
    \acs{eq} = \frac{\sum_{Z=0}^{2^d-1} \left| H_Z(C) - H_Z(P) \right|}{2^d}.
    \label{eq:eq}
\end{equation}

\begin{table}[htp]
\centering
\caption{Encryption Quality for \acp{ctc} video classes at five QP values}



\begin{tabular}{|c||c|c|c|c|c|c|}
\hline
 & \multicolumn{6}{c|}{Encryption Quality (EQ)} \\ \cline{2-7}
 & $QP$ & $QP$ & $QP$ & $QP$ & $QP$ & $EQ_{max} $\\
 & $17$ & $22$ & $27$ & $32$ & $37$ & \\
\hline
\hline
A1 	& 8 661	& 8 260	& 8 573	& 7 947	& 8 535	& 16 200 \\\hline
A2 	& 9 235	& 6 791	& 8 058	& 8 871	& 7 971	& 16 200 \\\hline
B 	& 1 620	& 1 805	& 1 923	& 1 920	& 1 911	& 4 050 \\\hline
C 	& 231	& 245	& 224	& 236	& 270	& 780 \\\hline
D 	& 61	    & 69	    & 81	    & 70	    & 83	    & 195 \\\hline
E 	& 682	& 748	& 1 057	& 1 150	& 1 096	& 1 800 \\\hline
F 	& 907	& 1 225	& 944	& 1 240	& 1 298	& 2 108 \\\hline
\hline
\textbf{Average} & \textbf{2640} & \textbf{2407} & \textbf{2603} & \textbf{2680} & \textbf{2652} & \textbf{5199}\\\hline
\end{tabular}
\vspace{-2mm}

\label{EQ_complete}
\vspace{-2em}
\end{table}

The higher \ac{eq} value is, the more secure is the selective encryption solution. Table~\ref{EQ_complete} presents the \ac{eq} values for all video classes at the five considered \acp{qp}. The presented values are the average \ac{eq} over encrypted frames and video sequences of each class. The \ac{eq} does not have a relative point for comparison. A derivation from~\eqref{eq:eq} is proposed to compute the upper bound value of the \ac{eq}~\cite{hamidouche2017real} as follows
\begin{equation}
    \acs{eq}_{max} = \frac{2 \, W \, H}{2^{d}},
    \label{eq:eqmax}
\end{equation}
where $H$ and $W$ are the video height and width, respectively and $d$ is the bit depth. The upper bound value of the \ac{eq} is reached when the histograms of the two frames $H_Z(C)$ and $H_Z(P)$ are not overlapping. 

The average \ac{eq} values are within the interval $[2407, \;  2680 ]$ in average with a theoretical average upper bound of 5199.  

The \ac{hevc} selective encryption solution proposed in~\cite{hamidouche2017real} achieved an \ac{eq} value for {\it Kimono} video sequence higher than 38.54\% of its maximum \ac{eq}, and an \ac{eq} value for {\it PeopleOnStreet} video higher than 40.92\% its maximum \ac{eq}. The proposed solution of different videos at different configuration ranges between 46.29\% and 51.56\% of the maximum \ac{eq} which confirms that the proposed solution has a high security level regarding the encryption quality metric.

\begin{figure*}[htpb]
    \subfloat[\label{fig:anchor_qp17_histo}$QP\ 17$]{\includegraphics[width=0.19\textwidth]{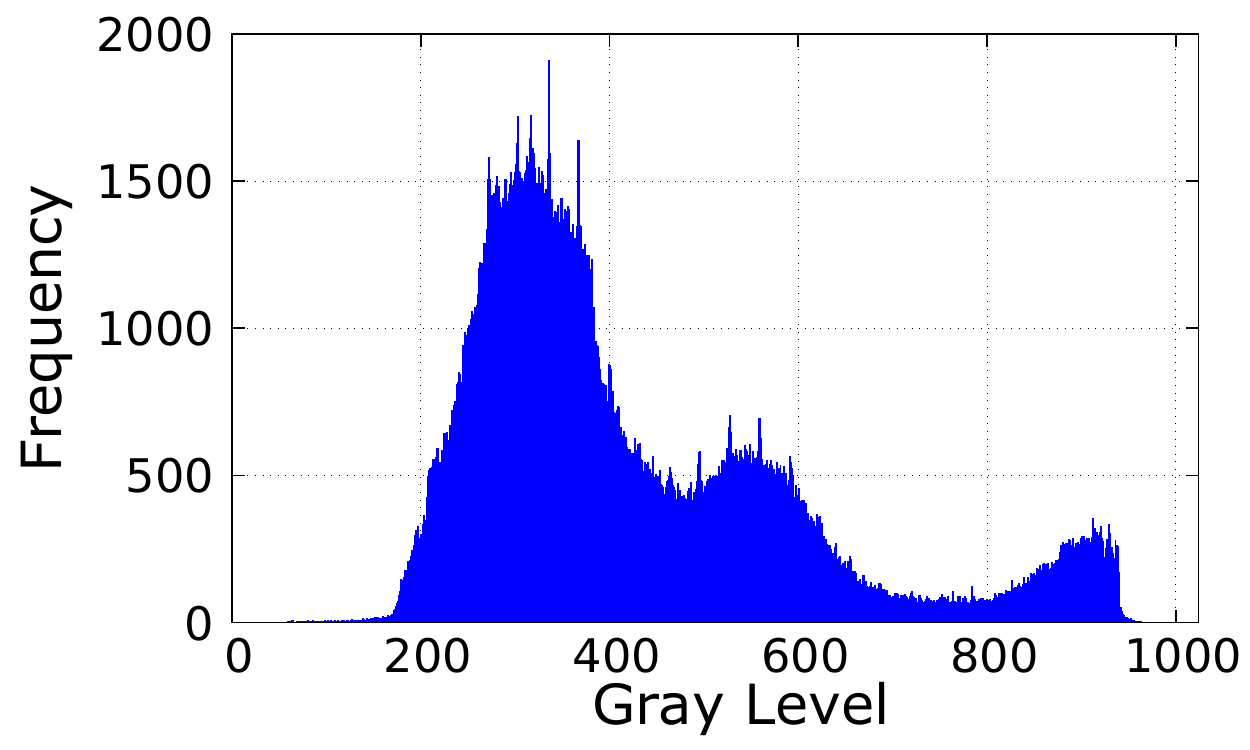}}     \hfill
    \subfloat[\label{fig:anchor_qp22_histo}$QP\ 22$]{\includegraphics[width=0.19\textwidth]{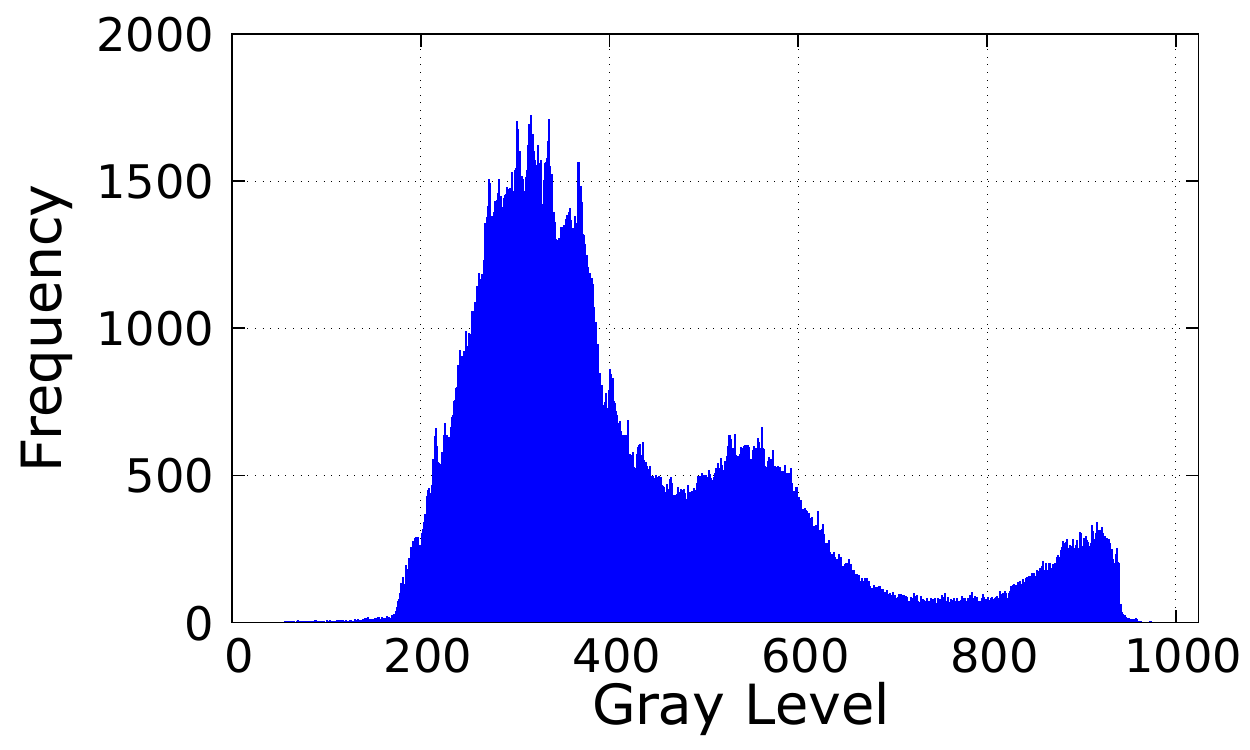}}     \hfill
    \subfloat[\label{fig:anchor_qp27_histo}$QP\ 27$]{\includegraphics[width=0.19\textwidth]{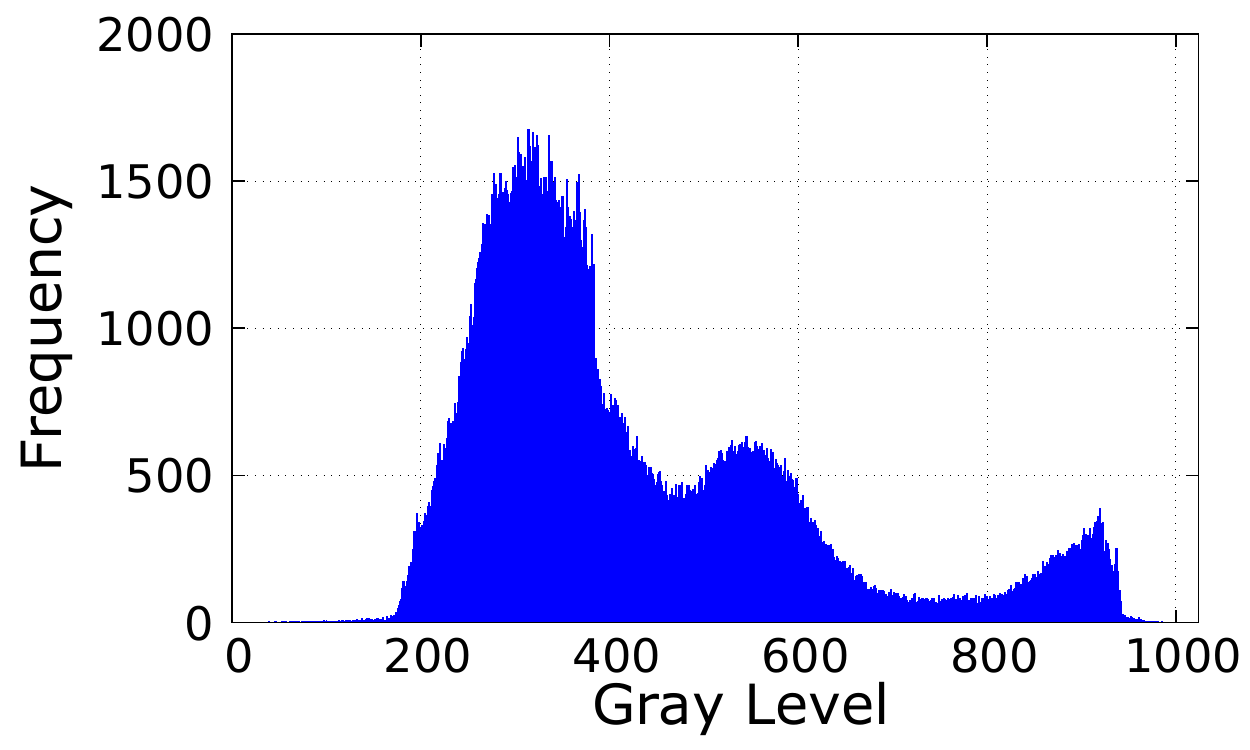}}     \hfill
    \subfloat[\label{fig:anchor_qp32_histo}$QP\ 32$]{\includegraphics[width=0.19\textwidth]{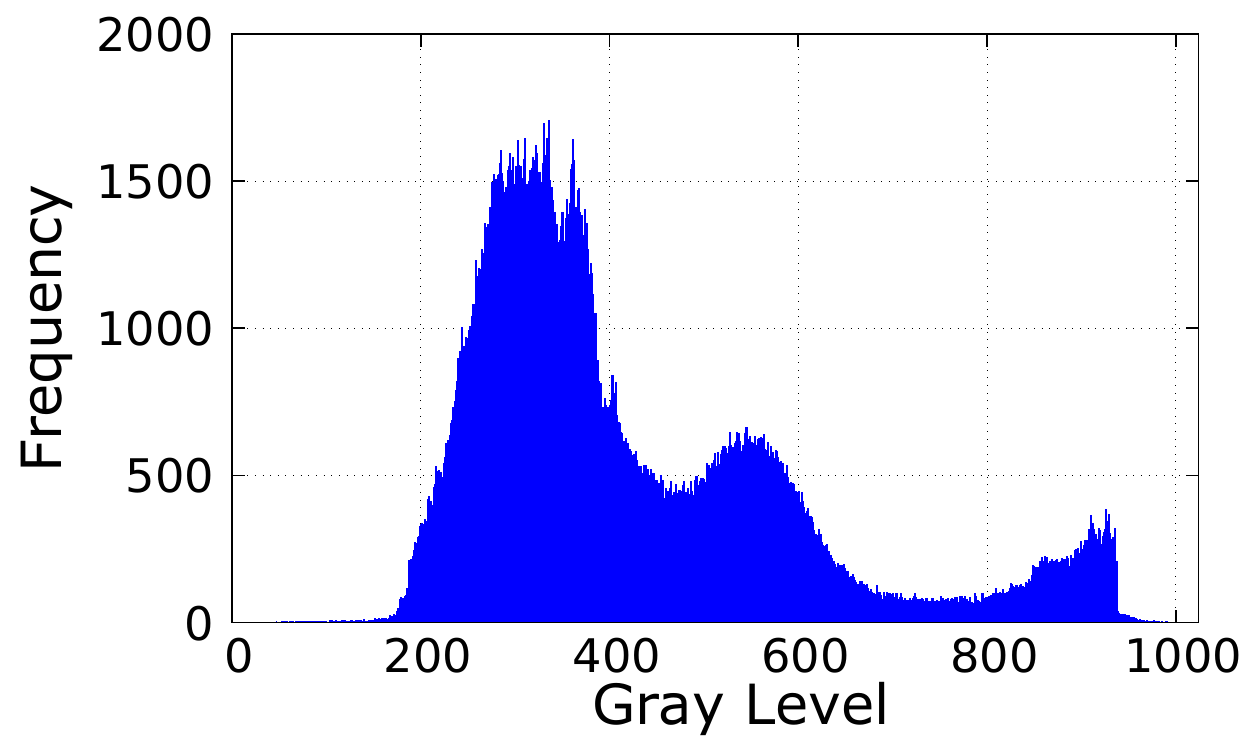}}     \hfill
    \subfloat[\label{fig:anchor_qp37_histo}$QP\ 37$]{\includegraphics[width=0.19\textwidth]{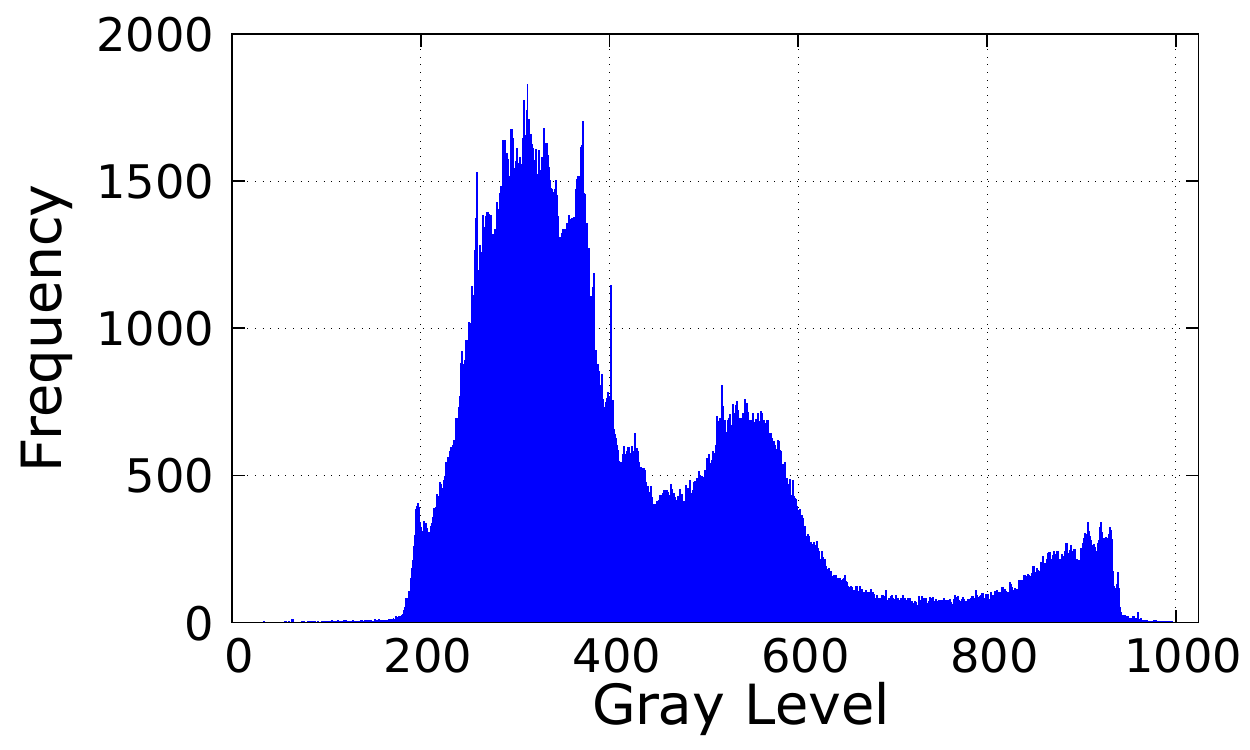}}     \hfill
    
    \subfloat[\label{fig:cipher_qp17_histo}$QP\ 17$]{\includegraphics[width=0.19\textwidth]{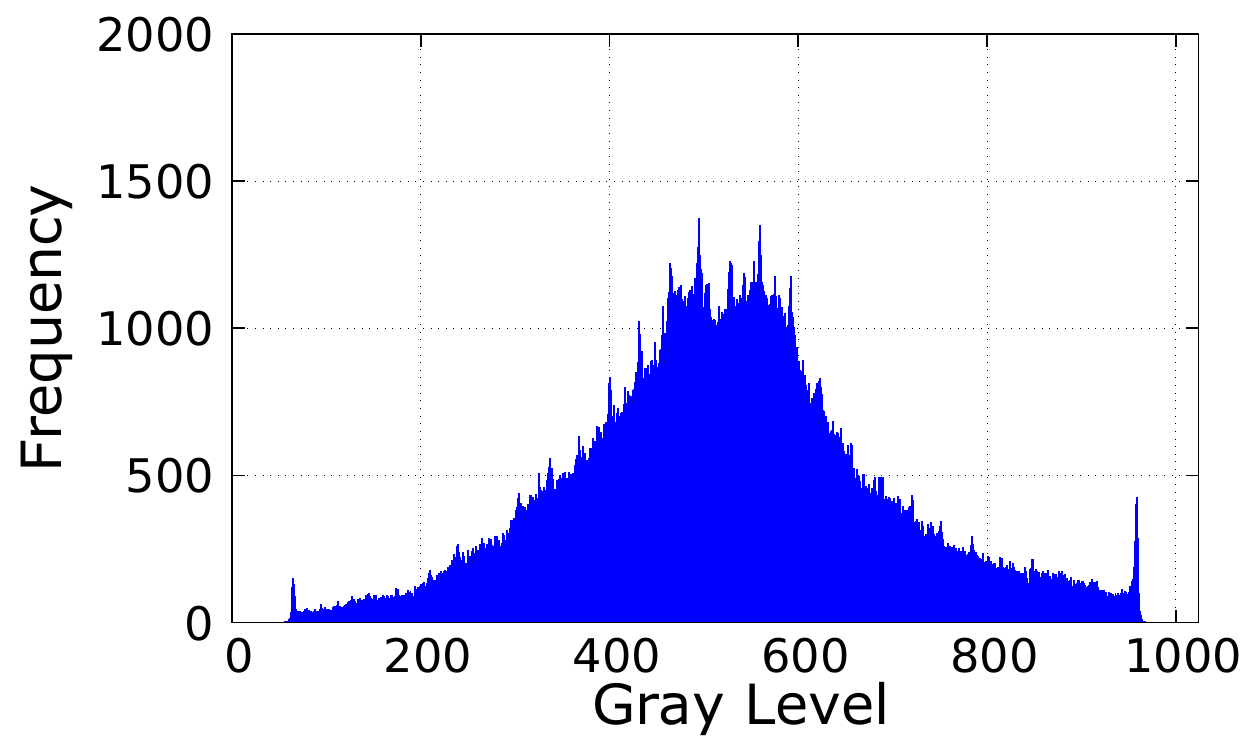}}     \hfill
    \subfloat[\label{fig:cipher_qp22_histo}$QP\ 22$]{\includegraphics[width=0.19\textwidth]{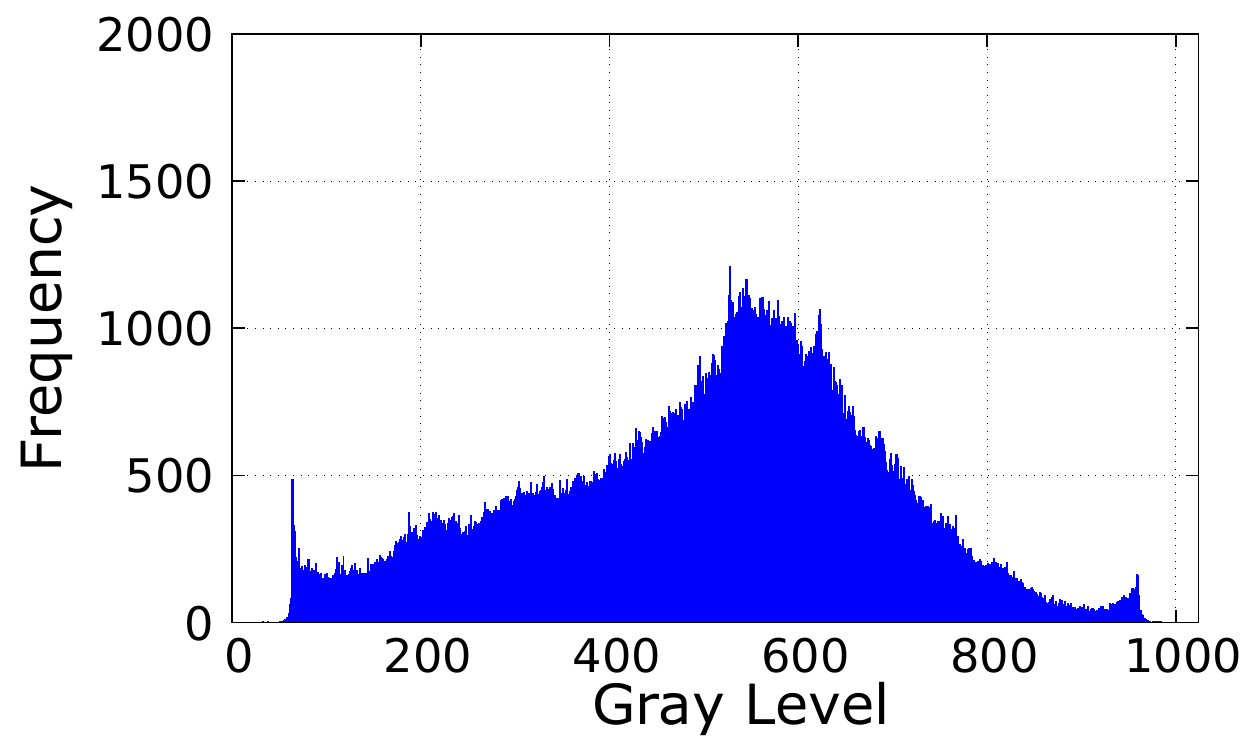}}     \hfill
    \subfloat[\label{fig:cipher_qp27_histo}$QP\ 27$]{\includegraphics[width=0.19\textwidth]{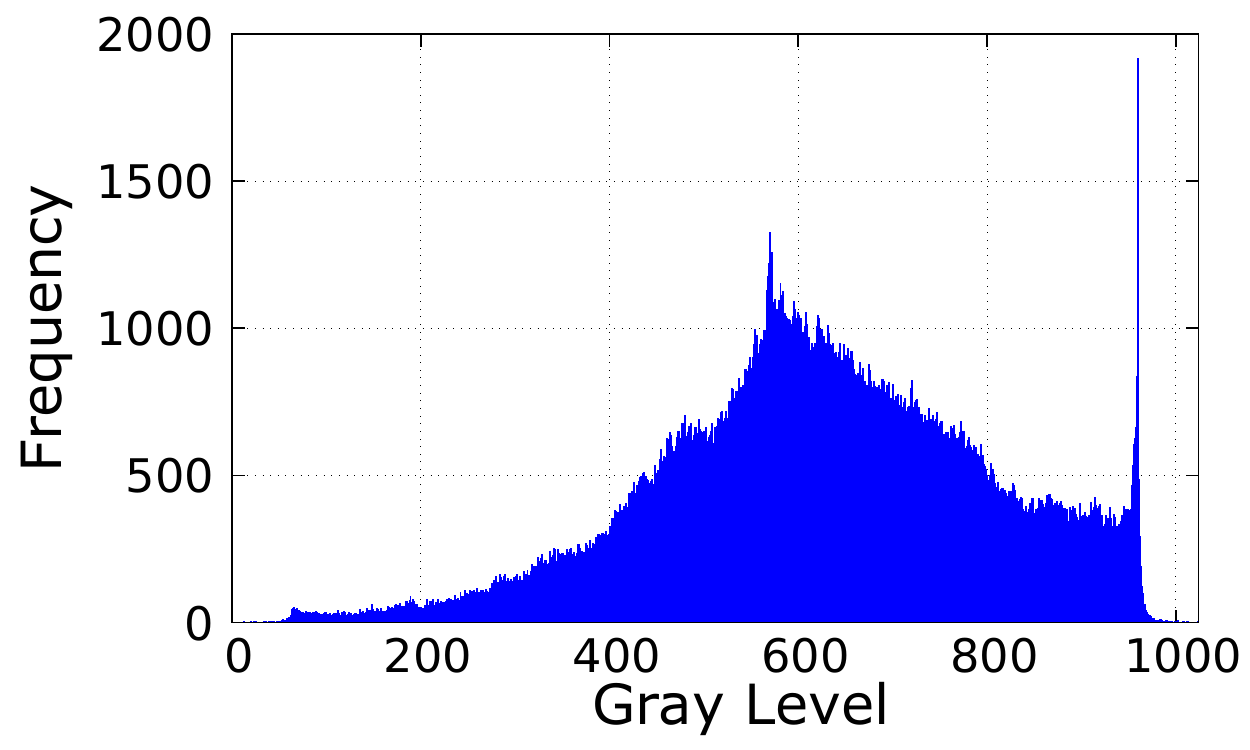}}     \hfill
    \subfloat[\label{fig:cipher_qp32_histo}$QP\ 32$]{\includegraphics[width=0.19\textwidth]{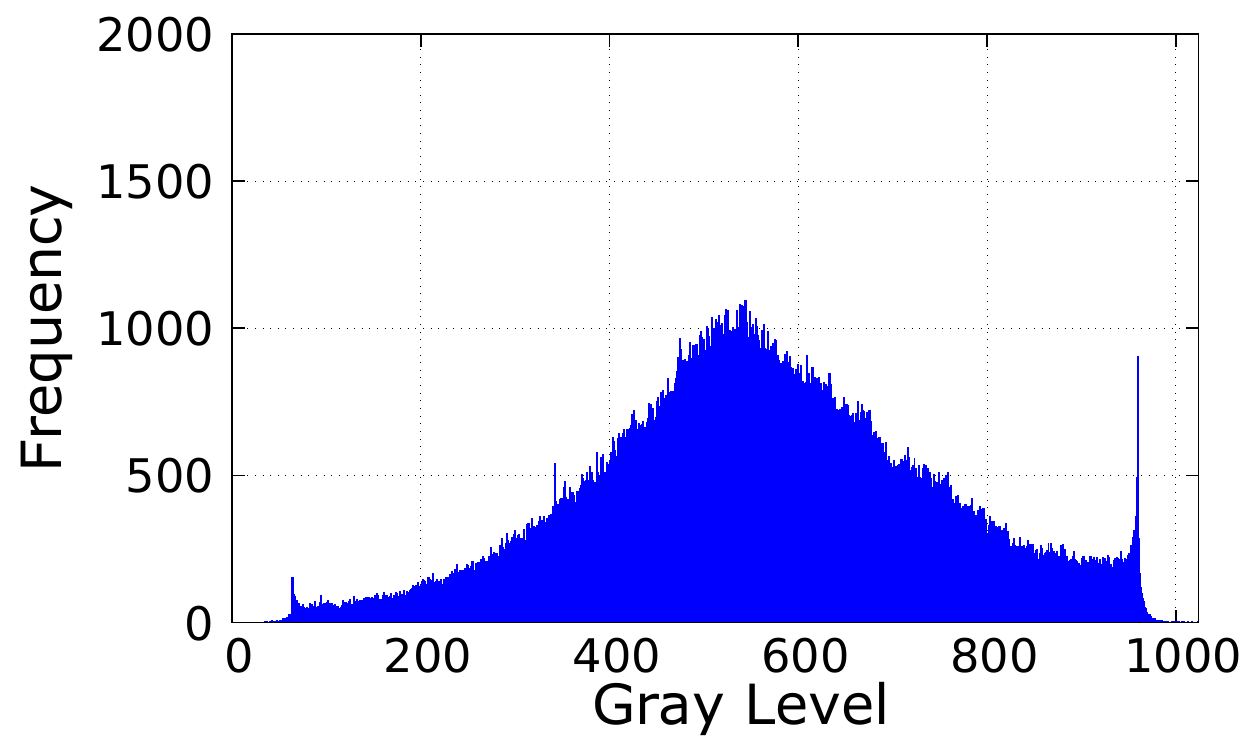}}     \hfill
    \subfloat[\label{fig:cipher_qp37_histo}$QP\ 37$]{\includegraphics[width=0.19\textwidth]{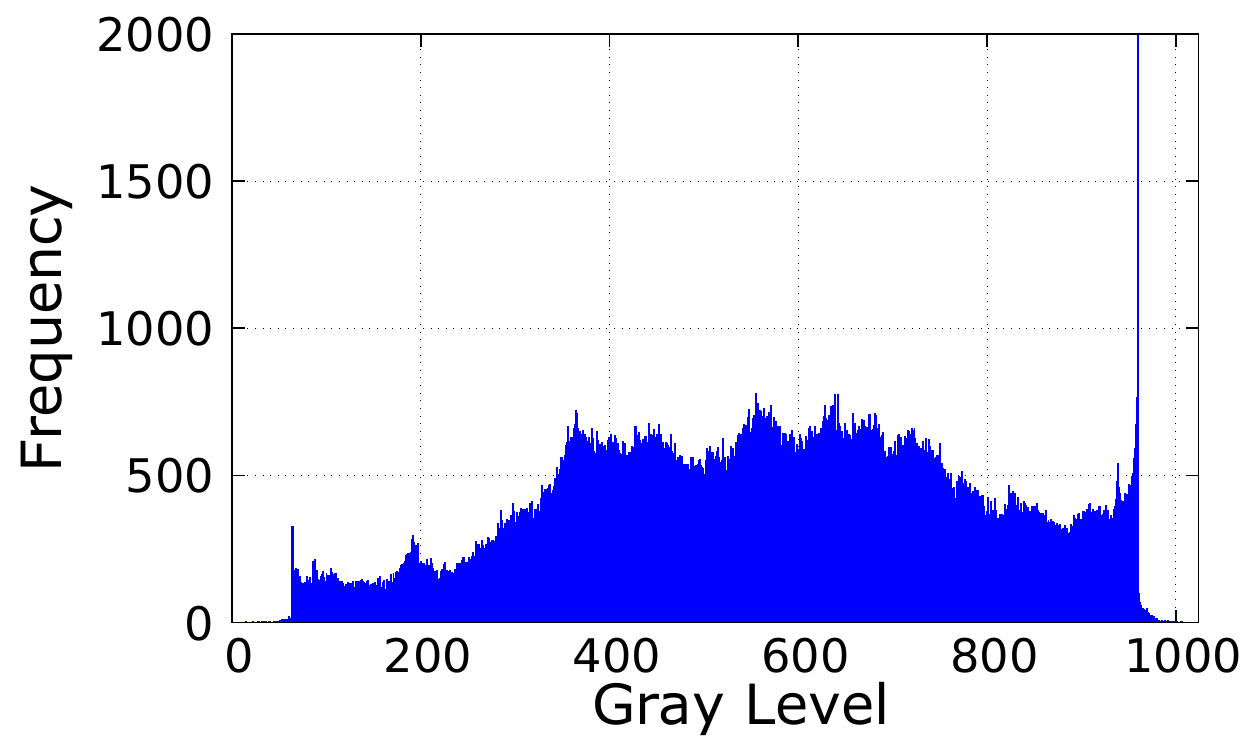}}     \hfill

    \caption{Histograms of frame \#10 computed for the anchor (\ref{fig:anchor_qp17_histo}- \ref{fig:anchor_qp37_histo}) and cipher (\ref{fig:cipher_qp17_histo}- \ref{fig:cipher_qp37_histo}) of {\it RaceHorsesC} video at five $QP$ values. The selective encryption solution significantly changes the pixels distributions as shown by the \ac{eq} metric.}
    \label{fig:snapshot_qp__histo}
    \vspace{-1em}
\end{figure*}

\subsubsection{\sc{Histogram analysis}}
Histogram of encrypted frame should be more uniform than original frame histogram in order to resist to statistical analysis based attacks~\cite{qumsieh2019joint,lewis1995data}. Fig.~\ref{fig:snapshot_qp__histo} illustrates the histograms of frame \#10 of {\it RaceHorcesC} video sequence before and after selective encryption at five considered \acp{qp}. 
The histograms of the encrypted frames is completely different from the original frame histogram. In fact, the proposed encryption solution changes the distribution of the decoded pixels toward different pattern which is close to uniform distribution especially at lower bitrate (ie. high \ac{qp}). \addcomment{We can also notice that in contrast to full encryption, it is difficult for constant bitrate and format compliant selective encryption to reach the uniform distribution of the histogram at all coding configurations and video contents.} 


\subsubsection{\sc{Edges and structural information protection}}
Edge detection enables assessing the ability of an encryption solution to hide the edge information in the encrypted frame. This section evaluates the ability of the proposed encryption solution to hide the edge in the encrypted video sequence. The \ac{edr} is computed by~\eqref{eq:edr}~\cite{taneja2011selective,taneja2011chaos}. 
\begin{equation}
    \acs{edr} = \frac{\sum_{i=1}^{H} \sum_{j=1}^{W} \left| P_{ED}(i,j) - C_{ED}(i,j) \right|}{\sum_{i=1}^{H} \sum_{j=1}^{W} \left| P_{ED}(i,j) + C_{ED}(i,j) \right|},
    \label{eq:edr}
\end{equation}
with $P_{ED}$ and $C_{ED}$ are the binary Laplacian of the decoded images with and without encryption, respectively. \Ac{edr} takes values in the interval  $[0, \,1]$, where a value close to 1 corresponds to a high edge hiding capability. 

Table~\ref{edr_table} presents the average values of \ac{edr} computed on the first 64 frames of the \acp{ctc} video sequences. The average \ac{edr} values are higher than 0.87 which shows the ability of the proposed selective encryption to hide edges and structural information in the encrypted frames. \addcomment{We can notice a lower \ac{edr} performance for class F video sequences. This class includes mainly screen content video sequences for which selective encryption is less effective to hide the structure of the edges.}  

\begin{table}[ht]
\centering
\caption{Average \ac{edr} for \acp{ctc} video classes at five QP values}


\begin{tabular}{|c||c|c|c|c|c|}
\hline
\multicolumn{1}{|c||}{}& \multicolumn{5}{c|}{EDR} \\ \cline{2-6}
\multicolumn{1}{|c||}{} & $QP\ 17$ & $QP\ 22$ & $QP\ 27$ & $QP\ 32$ & $QP\ 37$ \\ \cline{2-6}
\hline
\hline
A1        & 0.919 & 0.923 &	0.928 &	0.932 &	0.934 \\ \hline
A2        & 0.890 & 0.896 &	0.900 &	0.904 &	0.906 \\ \hline
B         & 0.890 & 0.893 &	0.896 &	0.899 &	0.904 \\ \hline
C         & 0.879 & 0.877 &	0.877 &	0.876 &	0.874 \\ \hline
D         & 0.876 & 0.887 &	0.890 &	0.880 &	0.881 \\ \hline
E         & 0.866 & 0.864 &	0.856 &	0.856 &	0.844 \\ \hline
F         & 0.813 & 0.780 &	0.762 &	0.741 &	0.762 \\ \hline
\hline

\textbf{Average}	  & \textbf{0.875} & \textbf{0.873} &	\textbf{0.871} &	\textbf{0.867} &	\textbf{0.871} \\ \hline
\end{tabular}

\label{edr_table}
\end{table}

\begin{figure*}
    \subfloat[\label{fig:anchor_qp17_edgedetection}$QP\ 17$]{\includegraphics[width=0.19\textwidth]{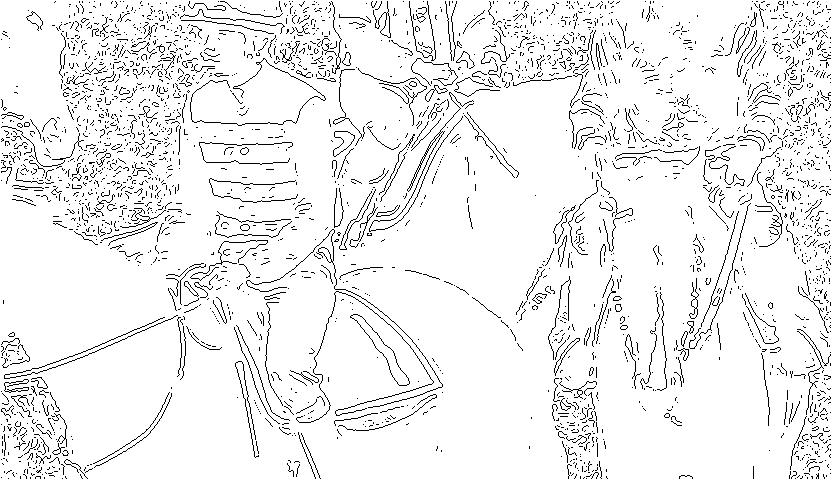}}     \hfill
    \subfloat[\label{fig:anchor_qp22_edgedetection}$QP\ 22$]{\includegraphics[width=0.19\textwidth]{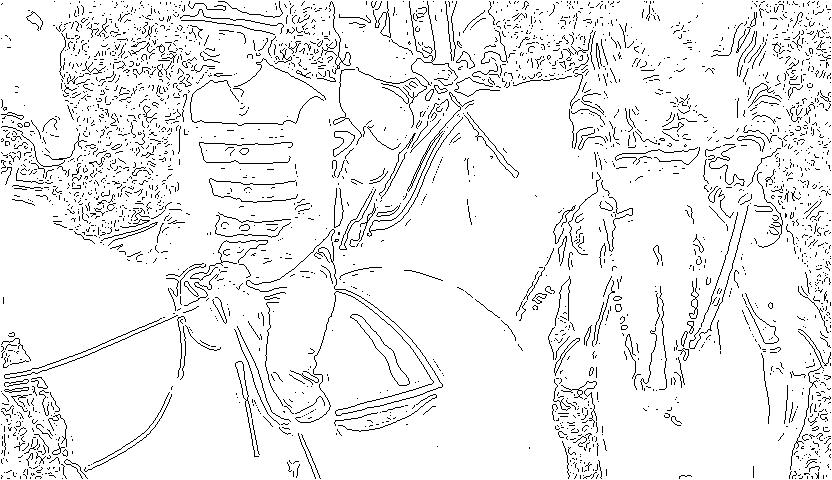}}     \hfill
    \subfloat[\label{fig:anchor_qp27_edgedetection}$QP\ 27$]{\includegraphics[width=0.19\textwidth]{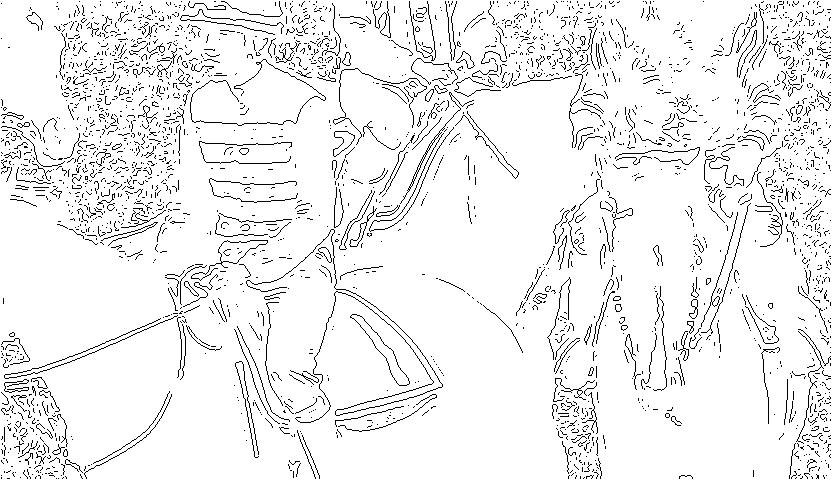}}     \hfill
    \subfloat[\label{fig:anchor_qp32_edgedetection}$QP\ 32$]{\includegraphics[width=0.19\textwidth]{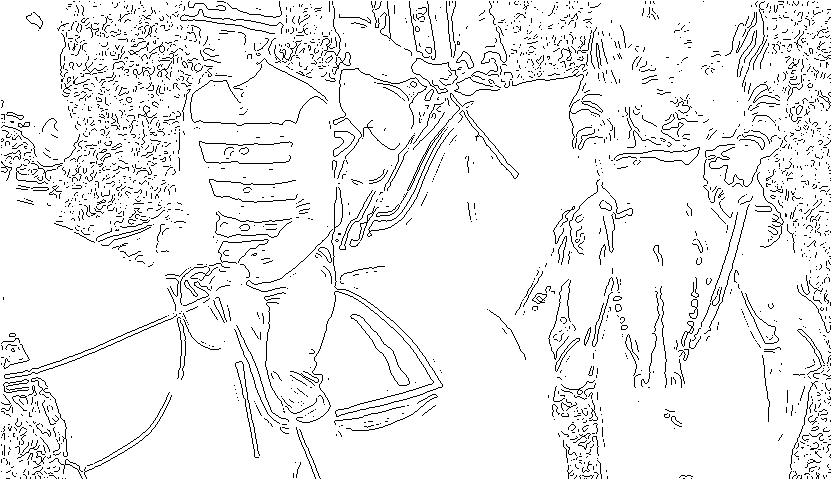}}     \hfill
    \subfloat[\label{fig:anchor_qp37_edgedetection}$QP\ 37$]{\includegraphics[width=0.19\textwidth]{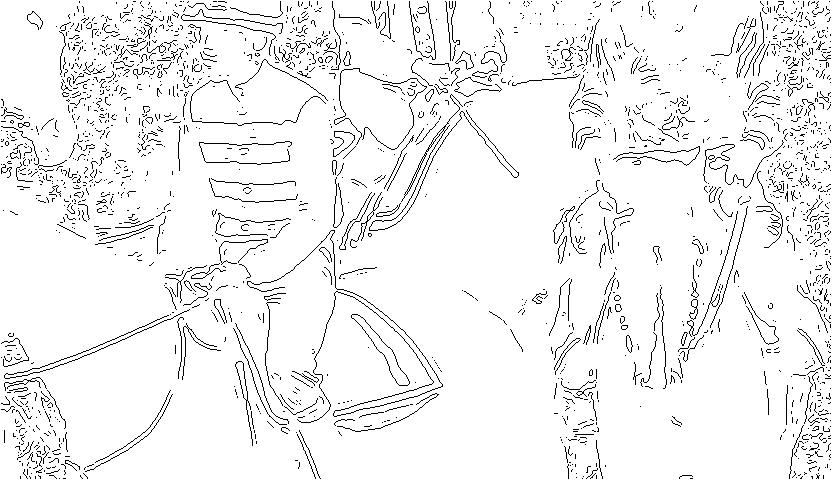}}     \hfill
    
    \subfloat[\label{fig:cipher_qp17_edgedetection}$QP\ 17$]{\includegraphics[width=0.19\textwidth]{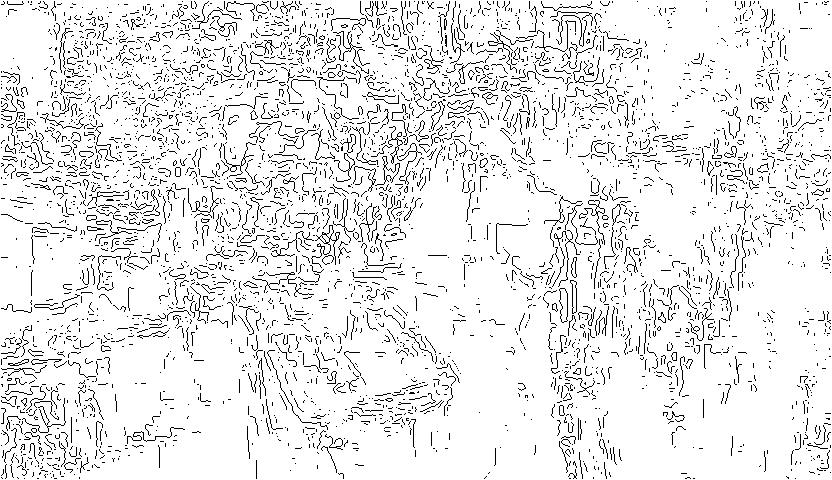}}     \hfill
    \subfloat[\label{fig:cipher_qp22_edgedetection}$QP\ 22$]{\includegraphics[width=0.19\textwidth]{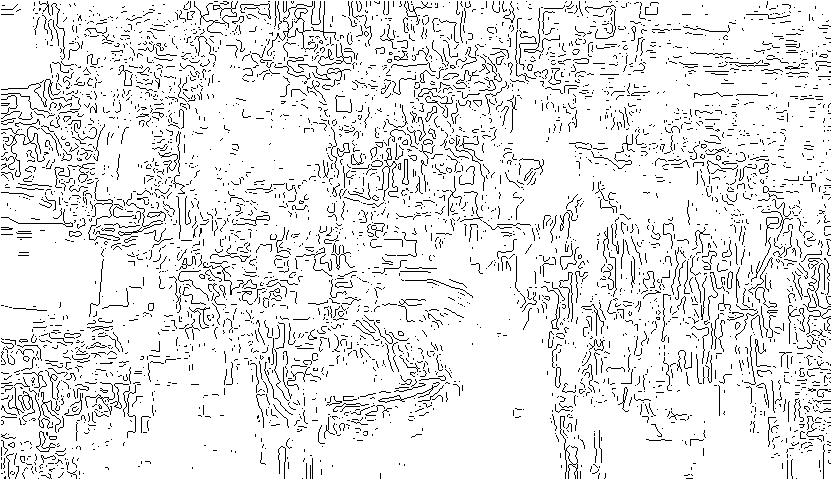}}     \hfill
    \subfloat[\label{fig:cipher_qp27_edgedetection}$QP\ 27$]{\includegraphics[width=0.19\textwidth]{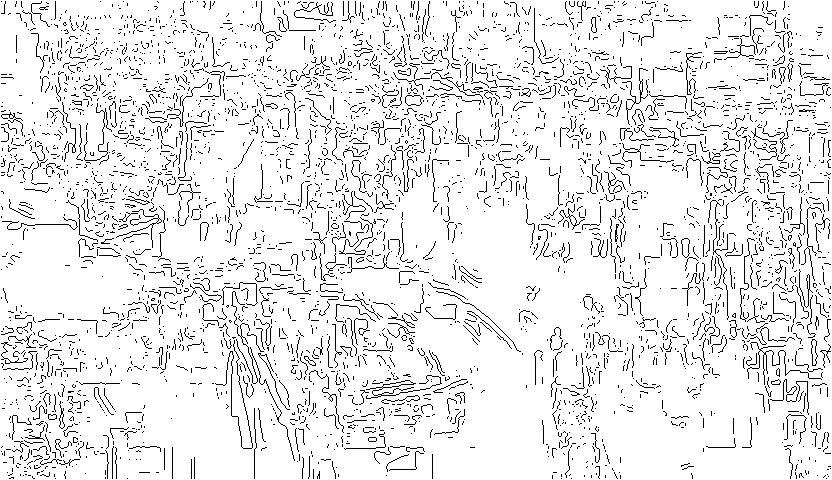}}     \hfill
    \subfloat[\label{fig:cipher_qp32_edgedetection}$QP\ 32$]{\includegraphics[width=0.19\textwidth]{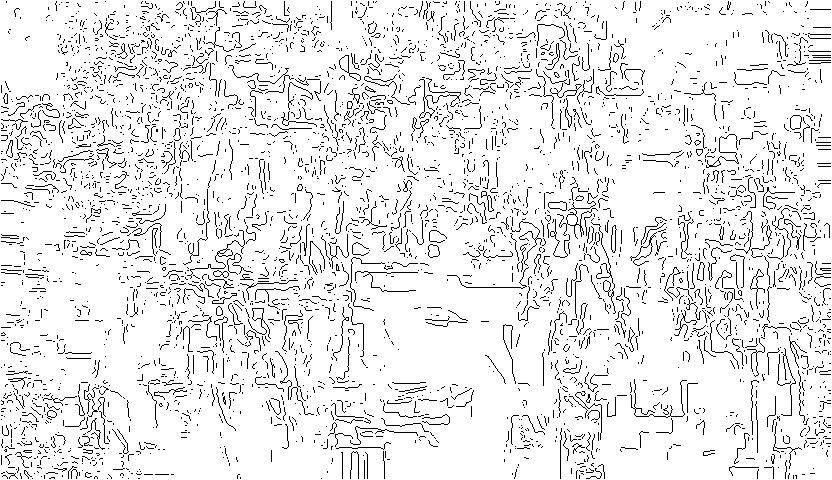}}     \hfill
    \subfloat[\label{fig:cipher_qp37_edgedetection}$QP\ 37$]{\includegraphics[width=0.19\textwidth]{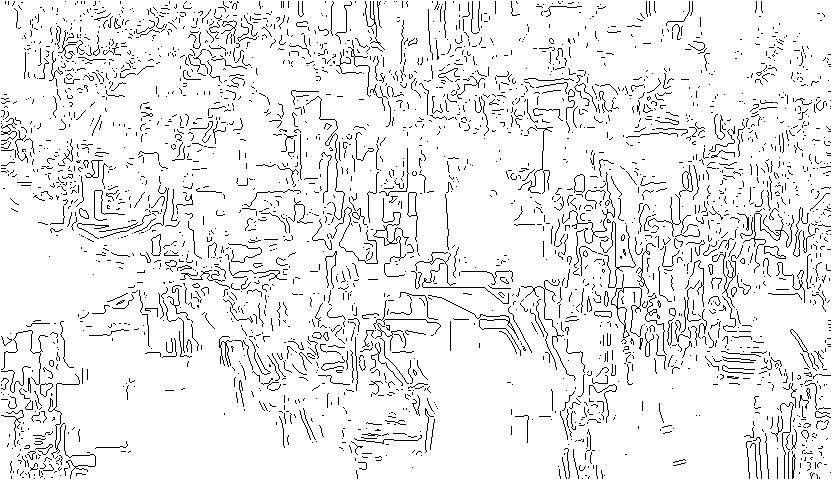}}     \hfill

    \caption{Edge Detection on Frame \#10 computed for the anchor (\ref{fig:anchor_qp17_edgedetection}-\ref{fig:anchor_qp37_edgedetection}) and ciphered (\ref{fig:cipher_qp17_edgedetection}- \ref{fig:cipher_qp37_edgedetection}) {\it RaceHorsesC} video at five \acp{qp}.}
    \label{fig:snapshot_qp_edgedetection}
    \vspace{-1em}
\end{figure*}

Fig.~\ref{fig:snapshot_qp_edgedetection} illustrates the edges of the decoded frame \#10 of {\it RaceHorsesC} sequence without encryption (first row) and with encryption at the second row for five different \acp{qp}. This figure clearly shows that the ciphered frames are noisy caused by high frequency structure introduced by the selective encryption. Therefore, structural information including edges in the ciphered frames are hidden and can be hardly explored by the \ac{edr} based attacks.


\subsubsection{\sc{Sensitivity to secret keys}}
Key sensitivity attacks are mainly based on the fact that the adversary tries to decipher the encrypted frames using a key close to the secret key used for encryption. The second adversary scenario is to guess the key if the encryption system provides information related to the used secrete key such as the sensitivity of the encryption regarding small change in the key. The proposed encryption algorithm should produce a completely different encrypted frame when a slight change (one bit change) on the used secrete key~\cite{pareek2013diffusion}.
Evaluation of the system robustness against key sensitivity attacks can be assessed using many existing tools such as \ac{uaci} and  \ac{npcr}~\cite{wu2011npcr, biham1991differential}. To compute these metrics, one random key is generated $K_1$ and a key with only one bit difference $K_2$ is created. The two keys are then used to cipher the same frame of width $W$ and height $H$ and a bit depth $d$. The result will create a ciphered frame $C_1$ using $K_1$, and $C_2$ using $K_2$. The \ac{uaci} and the \ac{npcr} are defined as follow
\begin{equation}
    \acs{uaci} = \frac{1}{H \, W \, 2^d} \, \sum_{i=1}^{H} \sum_{j=1}^{W} \left| C_1(i,j) - C_2(i,j) \right| \, 100 \%,
\end{equation}
\begin{equation}
    \acs{npcr} = \frac{1}{H \, W} \, \sum_{i=1}^{H} \sum_{j=1}^{W} D(i,j) \, 100 \%,
\end{equation}
with:
\begin{equation*}
    D(i,j) = \begin{cases} 
      0, & \text{if }C_1(i,j) = C_2(i,j) \\
      1, & \text{if }C_1(i,j) \neq C_2(i,j)
   \end{cases}
   .
\end{equation*}


\begin{table}[htp]
\caption{\ac{npcr} and \ac{uaci} with two secret keys with 1-bit-difference}


\begin{tabular}{|c||c|c||c|c||c|c|}
\hline
& \multicolumn{6}{c|}{$UACI$ and $NPCR$} \\ \cline{2-7}
& \multicolumn{2}{c||}{$QP \;  17$} & \multicolumn{2}{c||}{$QP \;  27$} & \multicolumn{2}{c|}{$QP \; 37$} \\ \cline{2-7}
& $UACI$ & $NPCR$ & $UACI$ & $NPCR$ & $UACI$ & $NPCR$  \\ \hline
\hline
A1      & 25.52  &  99.62    &  24.68  &  99.77    &  31.18  &  99.90 \\ \hline
A2      & 22.03  &  99.85   &  21.30  &  99.82    &  30.29  &  99.76 \\ \hline
B       & 24.48  &  99.83   &  24.27  &  99.82    &  27.04  &  99.60 \\ \hline
C       & 19.60  &  99.82   &  24.32  &  99.86    &  23.21  &  99.84 \\ \hline
D       & 22.18  &  99.85    &  23.12  &  99.83    &  23.51  &  99.69 \\ \hline
E       & 23.69  &  99.79    &  32.54  &  99.91    &  37.79  &  99.91 \\ \hline
F       & 26.84  &  99.52    &  26.96  &  99.33    &  22.65  &  98.76 \\ \hline
\hline
\textbf{Ave.} & \textbf{23.48}  &  \textbf{99.76}  &  \textbf{25.17}  & \textbf{ 99.76}  &    \textbf{27.32}  &  \textbf{99.61}\\\hline
\end{tabular}

\label{npcr_uaci_complete}
\end{table}
The optimal \ac{npcr} and \ac{uaci} values of a secure \addcomment{image} encryption scheme against key sensitivity attacks are 99.58\% and 33.46\%, respectively~\cite{maleki2008image}. 

Table~\ref{npcr_uaci_complete} presents the obtained \ac{npcr} and \ac{uaci} values on the \acp{ctc} for all video classes at three \acp{qp}. \addcomment{Here, it is important to note that the \ac{npcr} and \ac{uaci} results of the selective encryption should not be analysed as in  full image encryption. However,  the obtained values can give an indication on the ability of the selective encryption to resist key sensitive  and differential attacks. }
The average \ac{npcr} values at the three \acp{qp} for all classes are very close to the optimal value of a secure encryption scheme against key sensitivity attacks. Moreover, the average \ac{uaci} values lie in the interval $[19.60, \, 37.79]$ with an average value over all classes that converges to the optimal value of 33.64. This performance in terms of both \ac{npcr} and \ac{uaci} proves the robustness of the proposed selective encryption solution with regards the key sensitivity and differential attacks.
\subsubsection{\sc{Brute Force Attack}}
Brute force attack or exhaustive search attack performs testing all possible values of the used secret key in order to partially or completely break the cipher.~\cite{forouzan2007cryptography}
It is well-known, that any encryption algorithm with at least 128 bits as secret key is considered as resilient to brute force attack, which is the case for the used \ac{aes} algorithm. In selective encryption, the total number of tries to correctly guess the selected encrypted bits should be at least $2^{128}$ tries in order to resist to brute force attack~\cite{said2005measuring, choi2011privacy}. Our proposed selective encryption algorithm relies on \ac{aes} in counter mode as stream cipher with a secret key size of 128 bits. Moreover, the size of the encryption space is very large.

\subsubsection{\sc{Error Concealment Attack}}
The error concealment attack is a kind of attack based on guessing the encrypted bits based on some assumptions.

\begin{table*}[ht]
\centering
\renewcommand{\arraystretch}{1.1} 
	\caption{Replacement Attack, average \ac{psnr}, \ac{ssim} and \ac{vmaf} score on \acp{ctc}.}
	\label{tab:replacement}
	\begin{tabular}{|c||c|c|c|c|c|c|}
		\hline
		\multirow{2}{*}{\ac{qp}} & \multicolumn{2}{c|}{\ac{psnr}}  & \multicolumn{2}{c|}{\ac{ssim}}& \multicolumn{2}{c|}{\ac{vmaf}} \\ \cline{2-7}
		                    & Anchor & Replaced & Anchor & Replaced & Anchor & Repla.   \\ \hline
		                    \hline
		17                  & 44.85  & 5.87     & 1.00   & 0.29     & 99.21  & 6.77       \\ \hline
		22                  & 41.98  & 5.80     & 0.99   & 0.30     & 98.47  & 6.67       \\ \hline
		27                  & 39.42  & 5.90     & 0.99   & 0.31     & 95.78  & 7.16       \\ \hline
		32                  & 37.02  & 5.80     & 0.97   & 0.32     & 89.88  & 7.29       \\ \hline
		37                  & 34.54  & 5.92     & 0.95   & 0.32     & 80.21  & 6.93       \\ \hline
	\end{tabular}
	\vspace{-2mm}
\end{table*}

However, since the encryption space of a \ac{vvc} video is large, the only scenario that the adversary can follow is to try replacing all encrypted bits with the same value (zero or one) and decipher the modified encrypted frame \cite{stutz2009jpeg2000, dufaux2008scrambling}. In order to evaluate our proposed solution regarding error concealment attacks,  all encrypted bits are replaced by zero and then \ac{psnr}, \ac{ssim} and \ac{vmaf} are calculated again under the same \acp{ctc}. Table~\ref{tab:replacement} gives the average \ac{psnr}, \ac{ssim} and \ac{vmaf} scores of video sequences deciphered with replacement attack at the five \acp{qp}. The obtained quality scores are similar or even worst compared to encrypted video with \ac{aes} generator presented in Section~\ref{subsec:quality}. This confirms that the proposed selective encryption solution is robust against attacks based on replacement bits.

\subsection{\sc{Complexity Analysis}}
The  aim  of  this  section  is  to  assess  the  complexity  of the proposed encryption solution. The complexity overhead is  computed only for the decoder, since the encryption overhead is negligible with respect to the encoding time. 

\begin{table*}[ht]
\centering
\renewcommand{\arraystretch}{1.1} 
\caption{Deciphering time $\Delta_{SE}$ in second and deciphering overhead $CO_{SE}$ in \% on Intel i7-7700 processor at 3.6 GHz.}



\begin{tabular}{|c||c|c||c|c||c|c|}
\hline
& \multicolumn{2}{c||}{$QP=17$}  & \multicolumn{2}{c||}{$QP=27$} & \multicolumn{2}{c|}{$QP=37$} \\ \cline{2-7}
& $\Delta_{SE}$ & $CO_{SE}$ & $\Delta_{SE}$ & $CO_{SE}$ & $\Delta_{SE}$ & $CO_{SE}$ \\ \hline
\hline
A1     & 1.356  &  3.540    &  0.091  &  0.518  &  -0.013  & -0.147\\\hline
A2     & 2.514  &  5.605    &  0.156  &  0.674    &  -0.001  & -0.047\\\hline
B      & 0.473  &  4.579    &  0.039  &  0.976    &   0.001  &  0.041\\\hline
C      & 0.116  &  5.136    &  0.025  &  2.101    &   0.006  &  0.791\\\hline
D      & 0.026  &  4.476   &  0.005  &  1.384   &   0.001  &  0.450\\\hline
E      & 0.072  &  3.205    &  0.004  &  0.387    &  -0.002  & -0.198\\\hline
F      & 0.086  &  2.933   &  0.023  &  1.331   &   0.008  &  0.701\\\hline
\hline
\textbf{Average} & \textbf{0.581} & \textbf{4.236}  & \textbf{0.045} & \textbf{1.111} & \textbf{0.001} & \textbf{0.261}\\\hline
\end{tabular}
\vspace{-3mm}

\label{timetable}
\end{table*}

The average decoding run time is computed based on 100 decodings without deciphering ($DecT_{Ref}$) and with deciphering ($DecT_{SE}$). The deciphering run time $\Delta_{SE}$ is computed as a difference between decoding times with and without deciphering $\Delta_{SE} = DecT_{SE} - DecT_{Ref}$, while the percentage of deciphering complexity overhead $CO_{SE}$ is derived as follows 
\begin{equation}
CO_{SE} = \frac{\Delta_{SE}}{DecT_{Ref}} 100\%.
\end{equation}

Table~\ref{timetable} gives the deciphering time $\Delta_{SE}$ in second and the deciphering complexity overhead in percentage for all video classes at three \acp{qp}. The deciphering time does not exceed 3 seconds even for high bitrate and high resolution 4K videos of classes A and B. This corresponds to less than 6\% of the total decoding time. The average deciphering overhead remains lower than 4.23\% observed at high bitrate presenting more \acp{tc} to cipher. 

\section{Conclusion}
\label{sec:con}
In this paper a new selective encryption solution for the \ac{vvc} standard was proposed. This solution encrypts at the \ac{cabac} level a set of \ac{vvc} syntax elements in format-compliant and constant bitrate. The coding of the \acp{tc} in \ac{vvc} introduces several dependencies making constant bitrate encryption more challenging. We have proposed an original algorithm that analyses the coding dependencies of the \acp{tc} to determine the number and positions of encryptible bins for each coefficient. The proposed encryption solution was integrated in both encoder and decoder of the \ac{vvc} reference software \ac{vtm}~6.0. The quality of the encrypted video was assessed under the \ac{vvc} \acp{ctc} with three objective quality metrics including \ac{psnr}, \ac{ssim} and \ac{vmaf}. The low obtained quality scores clearly show the quality degradation enabled by the encryption. Security analysis was also conducted to asses the robustness against several attacks including statistical, key sensitivity and brute force attacks. Finally, the complexity overhead of the deciphering at the decoder side is estimated and remains lower than 6\% of the decoding time confirming the lightweight advantage of the proposed encryption solution. 




\bibliographystyle{IEEEtran}
\bibliography{arXiv_tam}

\begin{thebibliography}{10}
\providecommand{\url}[1]{#1}
\csname url@samestyle\endcsname
\providecommand{\newblock}{\relax}
\providecommand{\bibinfo}[2]{#2}
\providecommand{\BIBentrySTDinterwordspacing}{\spaceskip=0pt\relax}
\providecommand{\BIBentryALTinterwordstretchfactor}{4}
\providecommand{\BIBentryALTinterwordspacing}{\spaceskip=\fontdimen2\font plus
\BIBentryALTinterwordstretchfactor\fontdimen3\font minus
  \fontdimen4\font\relax}
\providecommand{\BIBforeignlanguage}[2]{{%
\expandafter\ifx\csname l@#1\endcsname\relax
\typeout{** WARNING: IEEEtran.bst: No hyphenation pattern has been}%
\typeout{** loaded for the language `#1'. Using the pattern for}%
\typeout{** the default language instead.}%
\else
\language=\csname l@#1\endcsname
\fi
#2}}
\providecommand{\BIBdecl}{\relax}
\BIBdecl

\bibitem{AES-FIPS}
``Specification for the advanced encryption standard (aes),'' Federal
  Information Process. Standards Publication 197, 2001.

\bibitem{liu2006efficient}
J.-L. Liu, ``Efficient selective encryption for jpeg 2000 images using private
  initial table,'' \emph{Pattern Recognition}, vol.~39, no.~8, pp. 1509--1517,
  2006.

\bibitem{jpegencry}
M.~V. Droogenbroeck, ``{Partial encryption of images for real-time
  applications},'' \emph{IEEE Signal Process. Symp.}, vol.~1, no.~2, pp.
  11--15, 2004.

\bibitem{JPEG200Survey}
E.~Dominik, S.~Thomas, and U.~Andreas, ``{A survey on JPEG2000 encryption},''
  \emph{IEEE Multimedia Systems}, vol.~15, no.~4, p. 243–270, 2009.

\bibitem{5733402}
Z.~{Shahid}, M.~{Chaumont}, and W.~{Puech}, ``Fast protection of h.264/avc by
  selective encryption of cavlc and cabac for i and p frames,'' \emph{IEEE
  Trans. Circuits Syst. Video Technol.}, vol.~21, no.~5, pp. 565--576, 2011.

\bibitem{4624035}
S.~{Park} and S.~{Shin}, ``Efficient selective encryption scheme for the
  h.264/scalable video coding(svc),'' in \emph{2008 Fourth Int. Conf. on
  Networked Computing and Advanced Information Management}, vol.~1, Sep. 2008,
  pp. 371--376.

\bibitem{shahid2013visual}
Z.~Shahid and W.~Puech, ``Visual protection of hevc video by selective
  encryption of cabac binstrings,'' \emph{IEEE Trans. Multimedia}, vol.~16,
  no.~1, pp. 24--36, 2013.

\bibitem{farajallah2015roi}
M.~Farajallah, W.~Hamidouche, O.~D{\'e}forges, and S.~El~Assad, ``Roi
  encryption for the hevc coded video contents,'' in \emph{2015 IEEE Int. Conf.
  on Image Process. (ICIP)}.\hskip 1em plus 0.5em minus 0.4em\relax IEEE, 2015,
  pp. 3096--3100.

\bibitem{7370952}
B.~{Boyadjis}, C.~{Bergeron}, B.~{Pesquet-Popescu}, and F.~{Dufaux}, ``Extended
  selective encryption of h.264/avc (cabac)- and hevc-encoded video streams,''
  \emph{IEEE Trans. Circuits Syst. Video Technol.}, vol.~27, no.~4, pp.
  892--906, April 2017.

\bibitem{hamidouche2017real}
W.~Hamidouche, M.~Farajallah, N.~Sidaty, S.~El~Assad, and O.~D{\'e}forges,
  ``Real-time selective video encryption based on the chaos system in scalable
  hevc extension,'' \emph{Signal Process.: Image Communication}, vol.~58, pp.
  73--86, 2017.

\bibitem{van2013encryption}
G.~Van~Wallendael, A.~Boho, J.~De~Cock, A.~Munteanu, and R.~Van~de Walle,
  ``Encryption for high efficiency video coding with video adaptation
  capabilities,'' \emph{IEEE Trans. Consum. Electron.}, vol.~59, no.~3, 2013.

\bibitem{memos2016encryption}
V.~A. Memos and K.~E. Psannis, ``Encryption algorithm for efficient
  transmission of hevc media,'' \emph{J. of Real-Time Image Process.}, vol.~12,
  no.~2, pp. 473--482, 2016.

\bibitem{long2018format}
M.~Long, F.~Peng, and X.~Gong, ``A format-compliant encryption for secure hevc
  video sharing in multimedia social network,'' \emph{Int. J. of Digit. Crime
  and Forensics (IJDCF)}, vol.~10, no.~2, pp. 23--39, 2018.

\bibitem{sallam2018efficient}
A.~I. Sallam, E.-S.~M. El-Rabaie, and O.~S. Faragallah, ``Efficient hevc
  selective stream encryption using chaotic logistic map,'' \emph{Multimedia
  Systems}, vol.~24, no.~4, pp. 419--437, 2018.

\bibitem{8954562}
N.~{Sidaty}, W.~{Hamidouche}, O.~{Déforges}, P.~{Philippe}, and J.~{Fournier},
  ``Compression performance of the versatile video coding: Hd and uhd visual
  quality monitoring,'' in \emph{2019 Picture Coding Symp. (PCS)}, 2019.

\bibitem{ITShannon}
C.~E. Shannon, ``{Communication theory of secrecy systems},''
  \emph{Declassified Report, Bell Syst. Tech. J.}, vol.~28, pp. 656--715, Dec
  1949.

\bibitem{EQmetric}
H.~E.~H. {Ahmed}, H.~M. {Kalash}, and O.~S.~F. {Allah}, ``Encryption efficiency
  analysis and security evaluation of rc6 block cipher for digit. images,'' in
  \emph{Int. Conf. on Electrical Engineering}, 2007.

\bibitem{taneja2011chaos}
N.~Taneja, B.~Raman, and I.~Gupta, ``Chaos based partial encryption of spiht
  compressed images,'' \emph{Int. J. of Wavelets, Multiresolution and
  Information Process.}, vol.~9, no.~02, pp. 317--331, 2011.

\bibitem{uacinpcr}
Y.~Wu, S.~Member, J.~P. Noonan, L.~Member, S.~Agaian, and S.~Member, ``Npcr and
  uaci randomness tests for image encryption,'' in \emph{Cyber Journals:
  Multidisciplinary Journals in Science and Technology, JSAT}, 2011.

\bibitem{shamir1979share}
A.~Shamir, ``How to share a secret,'' \emph{Commun. of the ACM}, vol.~22,
  no.~11, pp. 612--613, 1979.

\bibitem{vijayalakshmi2010efficient}
V.~Vijayalakshmi, L.~Varalakshmi, and G.~F. Sudha, ``Efficient encryption of
  intra and inter frames in mpeg video,'' in \emph{Int. Conf. on Network
  Security and Applications}.\hskip 1em plus 0.5em minus 0.4em\relax Springer,
  2010, pp. 93--104.

\bibitem{peng2019tunable}
F.~Peng, X.~Zhang, Z.-X. Lin, and M.~Long, ``A tunable selective encryption
  scheme for h. 265/hevc based on chroma ipm and coefficient scrambling,''
  \emph{IEEE Trans. Circuits Syst. Video Technol.}, 2019.

\bibitem{xudata}
D.~Xu, ``Data hiding in partially encrypted hevc video,'' \emph{ETRI J.}, 2020.

\bibitem{jvetVTM6doc}
S.~K. J.~Chen, Y.~Ye, ``{Algorithm description for Versatile Video Coding and
  Test Model 10 (VTM 10)},'' JVET, Tech. Rep. S2002, July 2020.

\bibitem{6317157}
V.~{Sze} and M.~{Budagavi}, ``High throughput cabac entropy coding in hevc,''
  \emph{IEEE Trans. Circuits Syst. Video Technol.}, vol.~22, no.~12, pp.
  1778--1791, 2012.

\bibitem{lipmaa2000comments}
H.~Lipmaa, P.~Rogaway, and D.~Wagner, ``Comments to nist concerning aes modes
  of operations: Ctr-mode encryption,'' in \emph{National Institute of
  Standards and Technologies}.\hskip 1em plus 0.5em minus 0.4em\relax Citeseer,
  2000.

\bibitem{boesgaard_rabbit_2008}
M.~Boesgaard, M.~Vesterager, and E.~Zenner, ``\BIBforeignlanguage{en}{The
  {Rabbit} {Stream} {Cipher}},'' in \emph{\BIBforeignlanguage{en}{New {Stream}
  {Cipher} {Designs}}}, ser. Lecture {Notes} in {Computer} {Science}.\hskip 1em
  plus 0.5em minus 0.4em\relax Springer, Berlin, Heidelberg, 2008, pp. 69--83.

\bibitem{gautier:hal-02184571}
G.~Gautier, M.~Le~Glatin, S.~El~Assad, W.~Hamidouche, O.~D{\'e}forges,
  S.~Guilley, and A.~Facon, ``{Hardware Implementation of Lightweight
  Chaos-Based Stream Cipher},'' in \emph{{Int. Conf. on Cyber-Technologies and
  Cyber-Systems}}, ser. CYBER 2019, Porto, Portugal, Sep. 2019, pp. 1--5.

\bibitem{wu_stream_2008}
H.~Wu, ``\BIBforeignlanguage{en}{The {Stream} {Cipher} {HC}-128},'' in
  \emph{\BIBforeignlanguage{en}{New {Stream} {Cipher} {Designs}}}, ser. Lecture
  {Notes} in {Comput.} {Sci.}\hskip 1em plus 0.5em minus 0.4em\relax Springer,
  Berlin, Heidelberg, 2008.

\bibitem{VtmGitlab}
\BIBentryALTinterwordspacing
``Git repository of the \ac{vtm}.'' [Online]. Available:
  \url{https://vcgit.hhi.fraunhofer.de/jvet/VVCSoftware_VTM}
\BIBentrySTDinterwordspacing

\bibitem{winkler2008evolution}
S.~Winkler and P.~Mohandas, ``The evolution of video quality measurement: From
  psnr to hybrid metrics,'' \emph{IEEE Trans. Broadcast.}, vol.~54, no.~3, pp.
  660--668, 2008.

\bibitem{wang2004image}
Z.~Wang, A.~C. Bovik, H.~R. Sheikh, and E.~P. Simoncelli, ``Image quality
  assessment: from error visibility to structural similarity,'' \emph{IEEE
  Trans. Image Process.}, vol.~13, no.~4, pp. 600--612, 2004.

\bibitem{vmaf}
R.~{Rassool}, ``Vmaf reproducibility: Validating a perceptual practical video
  quality metric,'' in \emph{2017 IEEE Int. Symp. on Broadband Multimedia
  Systems and Broadcasting (BMSB)}, June 2017, pp. 1--2.

\bibitem{qumsieh2019joint}
R.~Qumsieh, M.~Farajallah, and R.~Hamamreh, ``Joint block and stream cipher
  based on a modified skew tent map,'' \emph{Multimedia Tools and
  Applications}, vol.~78, no.~23, pp. 33\,527--33\,547, 2019.

\bibitem{lewis1995data}
M.~Lewis-Beck, \emph{Data analysis: An introduction}.\hskip 1em plus 0.5em
  minus 0.4em\relax Sage, 1995, no. 103.

\bibitem{taneja2011selective}
N.~Taneja, B.~Raman, and I.~Gupta, ``Selective image encryption in fractional
  wavelet domain,'' \emph{AEU-Int. J. of Electron. and Commun.}, vol.~65,
  no.~4, pp. 338--344, 2011.

\bibitem{pareek2013diffusion}
N.~K. Pareek, V.~Patidar, and K.~K. Sud, ``Diffusion--substitution based gray
  image encryption scheme,'' \emph{Digit. signal Process.}, vol.~23, no.~3, pp.
  894--901, 2013.

\bibitem{wu2011npcr}
Y.~Wu, J.~P. Noonan, S.~Agaian \emph{et~al.}, ``Npcr and uaci randomness tests
  for image encryption,'' \emph{Cyber journals: multidisciplinary journals in
  science and technology, JSAT}, vol.~1, no.~2, pp. 31--38, 2011.

\bibitem{biham1991differential}
E.~Biham and A.~Shamir, ``Differential cryptanalysis of des-like
  cryptosystems,'' \emph{J. of CRYPTOLOGY}, vol.~4, no.~1, pp. 3--72, 1991.

\bibitem{maleki2008image}
F.~Maleki, A.~Mohades, S.~M. Hashemi, and M.~E. Shiri, ``An image encryption
  system by cellular automata with memory,'' in \emph{2008 Third Int. Conf. on
  Availability, Reliability and Security}.\hskip 1em plus 0.5em minus
  0.4em\relax IEEE, 2008.

\bibitem{forouzan2007cryptography}
B.~A. Forouzan, \emph{Cryptography \& network security}.\hskip 1em plus 0.5em
  minus 0.4em\relax McGraw-Hill, Inc., 2007.

\bibitem{said2005measuring}
A.~Said, ``Measuring the strength of partial encryption schemes,'' in
  \emph{IEEE Int. Conf. on Image Process. 2005}, vol.~2.\hskip 1em plus 0.5em
  minus 0.4em\relax IEEE, 2005, pp. II--1126.

\bibitem{choi2011privacy}
S.~Choi, J.-W. Han, and H.~Cho, ``Privacy-preserving h. 264 video encryption
  scheme,'' \emph{ETRI Journal}, vol.~33, no.~6, pp. 935--944, 2011.

\bibitem{stutz2009jpeg2000}
T.~St{\"u}tz and A.~Uhl, ``On jpeg2000 error concealment attacks,'' in
  \emph{Pacific-Rim Symp. on Image and Video Technology}.\hskip 1em plus 0.5em
  minus 0.4em\relax Springer, 2009, pp. 851--861.

\bibitem{dufaux2008scrambling}
F.~Dufaux and T.~Ebrahimi, ``Scrambling for privacy protection in video
  surveillance systems,'' \emph{IEEE Trans. Circuits Syst. Video Technol.},
  vol.~18, no.~8, pp. 1168--1174, 2008.

\end{thebibliography}


\end{document}